\documentclass[fleqn,12pt]{article}
\usepackage[numbers,sort&compress]{natbib}
\usepackage[footnotesep=0.4in]{geometry}
\usepackage{graphicx}
\usepackage{amsmath}
\usepackage{bbding}
\usepackage{pifont}
\usepackage{amsfonts}
\usepackage{caption2}
\usepackage{float}
\usepackage{graphics}
\usepackage{subfigure}
\usepackage{float}
\usepackage{booktabs}
\usepackage{appendix}
\usepackage{ctex}
\makeatletter
\newcommand\figcaption{\def\@captype{\textbf{figure}}\caption}
\newcommand\tabcaption{\def\@captype{table}\caption}
\begin{document}
\date{}
\title{Breather transition dynamics, Peregrine combs/walls and modulation instability  in a variable-coefficient nonlinear Schr\"{o}dinger  equation with higher-order effects}
\author{Lei Wang$^{1}$$^,$\thanks{Corresponding
author:  50901924@ncepu.edu.cn},\,  Jian-Hui Zhang$^{2}$, Chong Liu$^{3}$, Min Li$^{1}$  and Feng-Hua Qi$^{4}$
\\{\em 1. Department of Mathematics and Physics, North China Electric}\\
{\em  Power University, Beijing 102206, P.\ R.\ China}
\\{\em 2. School of Energy Power and Mechanical Engneering,}\\ {\em North China Electric}
{\em  Power University, Beijing 102206, P.\ R.\ China}\\
{\em 3. School of Physics, Northwest University, Xi'an 710069, P.\ R.\ China}\\
{\em 4. School of Information, Beijing Wuzi University, }\\
{\em  Beijing 101149, P.\ R.\ China}
}
\maketitle

\newpage
\begin{abstract}
We study a variable-coefficient nonlinear Schr\"{o}dinger (vc-NLS) equation with higher-order effects. We show that the breather solution can be converted into four types of nonlinear waves on constant backgrounds including the multi-peak solitons, antidark soliton, periodic wave and W-shaped soliton. The transition condition requiring the group velocity dispersion (GVD) and third-order dispersion (TOD) to  scale linearly is obtained analytically. We display several kinds of elastic interactions between the transformed nonlinear waves. We discuss the dispersion management of multi-peak soliton, which indicates that the GVD coefficient controls the number of peaks of the wave while the TOD coefficient has compression effect. The gain or loss  has influence on the amplitudes of the multi-peak soliton. We further derive the breather multiple births by using multiple compression points of Akhmediev breathers in optical fiber systems with periodic dispersion. The number of ABs depends on the amplitude of the modulation but not on its wavelength, which affects their separation distance. In the limiting case, the breather multiple births reduce to the Peregrine combs. We discuss the effects of TOD coefficient on the spatiotemporal characteristics of Peregrine combs. When the amplitude of the modulation is equal to 1, the Peregrine comb is converted into a Peregrine wall that can be seen as intermediate state between rogue wave and W-shaped soliton. We finally find that  the modulational stability regions with zero growth rate coincide with the transition condition using rogue wave eigenvalues. Our results
could be useful for the experimental control and manipulation of the formation of generalized Peregrine
rogue waves in diverse physical systems modeled by vc-NLS equation with higher-order effects.

\vspace{3mm}\noindent\emph{\textbf{Keywords}}: Variable-coefficient nonlinear Schr\"{o}dinger equation,  breather transition dynamics,  Peregrine combs, Peregrine  walls, breather multiple births,  higher-order-effects management,  modulational instability

\vspace{3mm} \noindent\emph{\textbf{PACS numbers}}: 05.45.Yv, 42.81.Dp, 42.65.Hw\\

\end{abstract}
$\,$\vspace{2mm}
\newpage
\noindent\textbf{\Large{\uppercase\expandafter{1}. Introduction}}

Breathers have gotten a lot of attention due to their interactions and
energy exchange with a constant background~\cite{B-1}. Different from
the usual soliton dynamics,  their interactions can generate unique behaviors. Breathers can be classified into two kinds: Kuznetsov-Ma breathers (KMBs)~\cite{KMB} and Akhmediev breathers (ABs)~\cite{AB}. KMBs are  periodic in space and localized in time while ABs are periodic in time and localized in space.
Taking the period of both breather solutions to infinity produces a Peregrine soliton (PS) solution~\cite{PS}, which is localized both in space and time and  serves as a prototype of a rogue wave~\cite{RW}. This wave, which has a peak amplitude generally more than twice the significant wave height, appear from nowhere and disappear without a trace~\cite{TR}. It appears as a result of the modulation instability (MI)~\cite{MI-1,MI-2} of a weakly modulated plane wave. And more specifically, rogue-wave formation is related to a special kind of MI, namely, the baseband MI whose bandwidth includes arbitrarily small frequencies~\cite{FBO}. Experimentally, rogue waves have been observed in  optical fibers~\cite{L1}, water-wave tanks~\cite{L2}, and plasmas~\cite{L3}. In spite of many integrable equations have been shown to admit the rogue wave solutions, in theory, the standard nonlinear Schr\"{o}dinger (NLS) equation is seen as a basic model for the dynamics of rogue waves, both in water and optics.

In optical communications, there always exist some nonuniformities due to various factors that include the imperfection of manufacture,
variation in the lattice parameters of the fiber media and fluctuation of the fiber diameters~\cite{IN}. Those nonuniformities often lead to such effects as the fiber gain or loss,
phase modulation, and variable dispersion~\cite{IN1}. The inclusion of the variable coefficients into the NLS equations is currently an effective way to reflect the inhomogeneous effects of the nonlinear optical pulses~\cite{IN2}. On the other hand, if the dispersion and nonlinear effect in a mode-locked fiber laser is very strong,  the pulse parameters such as the width, chirp, phase and position will change significantly from their initial values~\cite{DM}. Thus, the concept of soliton dispersion management and soliton control in a fiber, which is used to address this problem,  has been recently proposed. The dispersion management of soliton is often modeled by the NLS equations with varying dispersion and nonlinear coefficients along with a gain or loss coefficient~\cite{DM2}. Compared with a conventional soliton, this type of soliton can not only be accelerated but also be amplified preserving its shape and elastic character, which makes it more suitable for diverse physical applications~\cite{DM}. Different from the nonlinear evolution equations with constant coefficients,  the rogue waves and breather in variable-coefficient ones can show some novel features such as the nonlinear tunneling effect, recurrence, annihilation and sustainment~\cite{NON,DAI,BEC1}, to name a few. Recent studies have also reported the breather evolution, amplification and compression, Talbot-like effects and composite rogue wave structures in some variable-coefficient nonlinear evolution equations~\cite{WL}. Moreover, Tiofack \emph{et al.} have demonstrated a novel multiple compression points structure in periodically modulated NLS equations, which is  termed as Peregrine comb~\cite{PCC}. Additionally, the studies of the control and manipulation of the rogue waves in variable-coefficient models may help to manage them experimentally in inhomogeneous optical fibers~\cite{FF,DAI} and Bose-Einstein condensates~\cite{BEC,BEC1}, and also provide a good fit between the theoretical analysis and real applications in future spatial observations and laboratory plasma experiments~\cite{GSM}.

The propagation of a picosecond optical pulse is usually described by the standard NLS equation. However, for the propagation of a subpicosecond or femtosecond pulse,
the higher-order effects such as the third order dispersion, self-steepening, and delayed nonlinear response should be taken into account~\cite{hirota}, which makes the modification of the NLS equation as a more accurate prototype of the wave evolution in the real world. These effects may add qualitatively certain new properties to the wave propagation phenomena, e.g., to breathers, rogue waves and MI.
Akhmediev \emph{et al.} have shown that a breather solution of the third- and fifth-order NLS equations can be converted into a nonpulsating soliton solution on a background, which does not exist in the standard NLS equation~\cite{AK1,AK2}. Wang \emph{{et al.}} have discovered that the breather solutions in the NLS equation with four-order dispersion and nonlinear terms  can be transformed into different types of nonlinear waves and the interactions between these waves are elastic~\cite{WLPRE}. Such transitions have also been reported in the higher-order coupled systems including the Hirota-Maxwell-Bloch (HMB) system~\cite{WLJP} and NLS and NLS-MB system with four-order effects~\cite{WLND}.
He \emph{et al.} have found that the higher-order terms control compression effects of the breather and rogue waves~\cite{HEPRE}.
With such higher-order perturbation terms as the TOD and delayed nonlinear response term, Liu \emph{et al.} have  found that the MI
growth rate shows a nonuniform distribution characteristic in the low perturbation frequency region, which opens up a stability region as the background frequency changes~\cite{LiuPRE}. They  have further exhibited an intriguing transition between bright-dark rogue waves and W-shaped-anti-W-shaped
solitons, which occurs as a result of the attenuation of MI growth rate to vanishing in the zerofrequency perturbation region~\cite{LiuAOP}.

In this paper, we consider a variable-coefficient NLS  (vc-NLS) equation with higher-order effects as follows~\cite{LL,LWJ-VCH,HJS-VCH}
\begin{equation}\label{vcHirota}
i\,q_{z}+\frac{d_{2}(z)}{2}q_{tt}+R(z)|q|^{2}q-i\,d_{3}(z)q_{ttt}-i\,6\,\gamma(z)|q|^{2}q_{t}-\frac{i}{2}\Gamma(z)q=0,
\end{equation}
with
 \begin{equation}
\begin{aligned}\label{cc}
\frac{\gamma(z)}{d_{3}(z)}=\frac{R(z)}{d_{2}(z)}\,, \qquad
 \Gamma=\frac{W[R(z),d_{2}(z)]}{d_{2}(z)R(z)}\,,  \qquad W[R(z),d_{2}(z)]=R\,d_{2z}-d_{2}\,R_{z},
\end{aligned}
\end{equation}
where $z$ is the propagation variable, $t$ is the retarded time in a moving frame with the group velocity, and $q(z, t)$ is the slowly varying envelope of the wave field.
The coefficients $d_{2}(z)$, $R(z)$,  $d_{3}(z)$,  $\gamma(z)$ and $\Gamma(z)$ represent the GVD effect, Kerr nonlinear effect,  TOD effect, the time-delay correlation to the cubic term and the gain or loss effect respectively. Liu \emph{et al.} have studied the two-soliton interactions of Eq.~(\ref{vcHirota}) analytically and numerically, and discussed the higher-order-effects management of soliton interactions~\cite{LWJ-VCH}. He \emph{et al.} have investigated the  control and manipulation of the rogue waves of  Eq.~(\ref{vcHirota})~\cite{HJS-VCH}.

Our goals here are twofold:  (1) the breather transition dynamics and nonlinear wave management; (2) the breather multiple births and the Peregrine combs/walls. We first present intriguing different kinds of nonlinear localized and periodic waves, including the multi-peak soliton,  W-shaped soliton and periodic wave. With the multi-peak soliton, for example, we discuss the effects of GVD coefficient, TOD coefficient and gain or loss coefficient. We further reveal the relation between such transition and MI characteristics. On the other hand, considering periodic modulation, we display three types of multiple compression points structures, the breather multiple births, Peregrine combs and Peregrine walls whose spatiotemporal characteristics are also analyzed analytically.

The arrangement of the paper is as follows: In Sec.~2, we will present different types of transformed nonlinear waves of Eq.~(\ref{vcHirota}), show their interactions,  and analyze the effects of variable coefficients. In addition, the transition condition will be given analytically. The relation between the MI growth rate and  transition condition will be revealed in Sec.~3.  The Peregrine comb/wall structures as well as their spatiotemporal characteristics will be studied in Sec.~4.   Finally, Sec.~5 will be the conclusions of this paper.

\vspace{5mm}
\noindent\textbf{\Large{\uppercase\expandafter{2}. Breather transition dynamics}}

\textbf{A. Breather-to-soliton transitions}

In this section, we mainly study the breather transition dynamics  for Eq.~(\ref{vcHirota}). By virtue of the Darboux transformation (see Appendix), the first-order breather solution of Eq.~(\ref{vcHirota}) can be derived as
\begin{equation}\label{EXP-B1}
\begin{aligned}
q_{B}^{[1]}=c(z)\,\bigg(1+2\,\beta\,\frac{G_{B}^{[1]}+i\,H_{B}^{[1]}}{D_{B}^{[1]}}\bigg)\,e^{i\,\rho}\,,
\end{aligned}
\end{equation}
 with
 \begin{equation}
\begin{aligned}
 &\rho=m(z)+n\,t\,,\quad m(z)=\int\Big(-\frac{1}{2}(n^{2}-2)d_{2}(z)-n(n^{2}-6)d_{3}(z)\Big)dz\,,\nonumber \\
&G_{B}^{[1]}=k_{1}k_{2}\cos(z\,V_{H}+t\,h_{R})\,\cosh(2\,\chi_{I})-\cosh(z\,V_{T}+t\,h_{I})\,\sin(2\,\chi_{R})\,,\nonumber \\
& H_{B}^{[1]}=\cos(2\,\chi_{R})\,\sinh(z\,V_{T}+t\,h_{I})+k_{1}k_{2}\sin(z\,V_{H}+t\,h_{R})\,\sinh(2\,\chi_{I})\,,\nonumber\\
& D_{B}^{[1]}=\cosh(z\,V_{T}+t\,h_{I})\,\cosh(2\,\chi_{I})-k_{1}k_{2}\cos(z\,V_{H}+t\,h_{R})\,\sin(2\,\chi_{R})\,,\nonumber\\
& h=2\,\sqrt{1+\Big(\lambda+\frac{n}{2}\Big)^{2}}=h_{R}+i\,h_{I}\,,\,\,k_{1}=1\,,\,\,k_{2}=\pm1\,,\nonumber \\
&\varpi=(t-\frac{1}{2}\int\Big((n-2\,\lambda)d_{2}(z)+2(-2+n^{2}-2\,n\,\lambda+4\,\lambda^{2})d_{3}(z)\Big)dz)\frac{h}{2}\notag \nonumber \\
&\quad=(t+\varpi_R(z)+i\, \varpi_I(z))\frac{h}{2}\,,\notag \nonumber\\
&\chi=\frac{1}{2}\arccos{\frac{h}{2}}\,,\quad c(z)=\sqrt{\frac{d_{2}(z)}{R(z)}}=\sqrt{\frac{d_{3}(z)}{\gamma(z)}}=\exp(\frac{1}{2}\Gamma (z))\,,\quad \lambda=\alpha+\beta\,i\,,\\
& V_{T}=2(\varpi_{R}(z)\,h_{I}+\varpi_{I}(z)\,h_{R})\,,
\quad V_{H}=2(\varpi_{R}(z)\,h_{R}-\varpi_{I}(z)\,h_{I})\,.\nonumber
\end{aligned}
 \end{equation}
From the above expression, one can find that the breather solution~(\ref{EXP-B1}) is composed of the hyperbolic functions $\sinh (z\,V_{T}+t\,h_{I})$ ($\cosh (z\,V_{T}+t\,h_{I})$) and trigonometric functions $\sin (z\,V_{H}+t\,h_{R})$ ($\cos (z\,V_{H}+t\,h_{R})$), where $\varpi_{R}(z)+\frac{\varpi_{I}(z)h_{R}}{h_{I}}$ and $\varpi_{R}(z)-\frac{\varpi_{I}(z)h_{I}}{h_{R}}$ are the corresponding velocities. The hyperbolic functions and trigonometric functions describe the localization and  periodicity of the transverse distribution $t$ of those waves, respectively. The
nonlinear wave described by the solution~(\ref{EXP-B1}) could be seen as a nonlinear superposition of a soliton and a periodic wave. The period of the breather along $t$-coordinate axis is determined by $\frac{\pi}{h_{R}}$ which is related to the eigenvalue $\lambda=\alpha+\beta\,i$. Compared with the Hirota equation with constant coefficients, the breather solution~(\ref{EXP-B1}) of the inhomogeneous Hirota equation~(\ref{vcHirota}) includes the variable dispersion ($d_{2}(z)$), nonlinearity ($R(z)$), higher-order effects ($d_{3}(z)$ and $\gamma(z)$) and gain or loss ($\Gamma(z)$). More specifically, the intensity of the breather is controlled by $c(z)=\sqrt{\frac{d_{2}(z)}{R(z)}}=\sqrt{\frac{d_{3}(z)}{\gamma(z)}}=\exp(\frac{1}{2}\Gamma (z))$, which means that we can manipulate the intensity by adjusting the gain or loss coefficient or the ratio of dispersion and nonlinearity. In addition, it is observed that $V_{T}$ and $V_{H}$ are associated with the GVD effect $d_{2}(z)$ and  TOD effect $d_{3}(z)$ that affect the velocity of the breather.  Thus, we will study the dynamics of the breather described by the solution~(\ref{EXP-B1}) depending on the above parameters.

Firstly, we  show that the breather solution~(\ref{EXP-B1}) can be converted into four types of nonlinear waves  depending on the values of the velocity difference $\frac{\varpi_{I}(z)(h_{R}^{2}+h_{I}^{2})}{h_{R}h_{I}}$.
When  $\varpi_{I}(z)(\frac{h_{R}^{2}+h_{I}^{2}}{h_{R}h_{I}})\neq0$ (or $\varpi_{I}(z)\neq0$), the solution~(\ref{EXP-B1}) characterizes the localized waves with breathing behavior on constant backgrounds (i.e., the breathers and rogue waves).  Further, if $\alpha=-\frac{n}{2}$, we have the ABs with $|\beta|<1$, the KMBs with $|\beta|>1$, and the PS with $|\beta|=1$. Those solutions have been obtained  via the similarity transformation~\cite{HJS-VCH}.

Conversely, if  $\varpi_{I}(z)=0$, the soliton and  periodic wave in the solution~(\ref{EXP-B1}) have the same velocity $\varpi_{R}(z)$. Further, we find that the case $\varpi_{I}(z)=0$ is equivalent to the following condition
 \begin{equation}\label{KZFC1}
    \begin{aligned}
&\frac{V_{T}}{h_{I}}=\frac{V_{H}}{h_{R}}\,,
  \end{aligned}
 \end{equation}
 i.e.,
\begin{equation}\label{KZFC}
n=4\,\alpha-\frac{d_{2}(z)}{2\,d_{3}(z)}\,.
\end{equation}
The condition~(\ref{KZFC1})  indicates that the extrema of trigonometric and hyperbolic functions in the solution~(\ref{EXP-B1}) is located along the same straight lines in the $(z,t)$-plane, which leads to the transformation of the breather into different types of nonlinear waves on constant backgrounds. The equivalent form of the condition~(\ref{KZFC1}), namely Eq.~(\ref{KZFC}), involves four parameters: the frequency of plane wave $n$,  the real part of the eigenvalue $\alpha$, and the GVD effect $d_{2}(z)$ and  TOD effect $d_{3}(z)$. When $d_{2}(z)$ is proportional to $d_{3}(z)$ ($d_{2}(z)=k\,d_{3}(z)$, $k\neq0$), Eq.~(\ref{KZFC}) has solutions. However, if $d_{2}(z)=k(z)\,d_{3}(z)$ ($k(z)$ is  a function of  $z$), Eq.~(\ref{KZFC}) has no solution. This means that the constraint $d_{2}(z)=k\,d_{3}(z)$ is the necessary conditions for the existence of transformed nonlinear waves.  For a fixed $n$, increasing the value of $\frac{d_{2}(z)}{d_{3}(z)}$ results in an decrease of $\alpha$.   This is plotted in \textbf{Fig.~1}.
\begin{figure}[H]\centering
\quad
\includegraphics[width=220 bp]{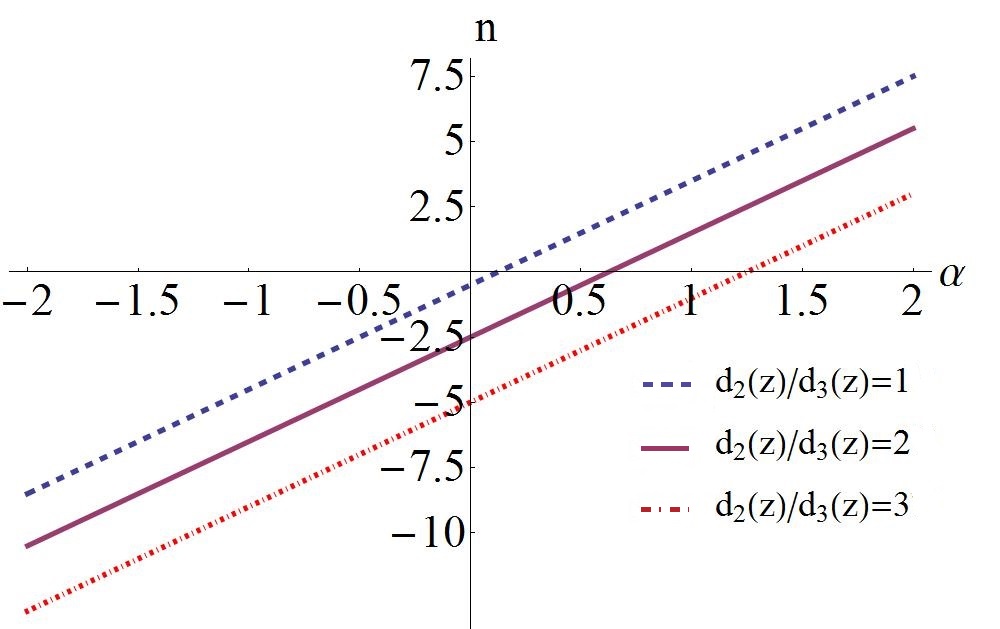}
\caption{\footnotesize Solutions of Eq.~(\ref{KZFC}) on a plane of ($n, \alpha$).}
\end{figure}
Under the transition condition~(\ref{KZFC}), we exhibit four kinds of transformed nonlinear waves on constant backgrounds including the multi-peak solitons [\textbf{Figs.~2(a)} and \textbf{2(b)}], antidark soliton (\textbf{Fig.~2(c)}), periodic wave [\textbf{Fig.~2(d)}] and W-shaped soliton [\textbf{Fig.~2(e)}]. These types of nonlinear waves have been found not only in the scalar equations including Hirota equation~\cite{AK1}, four-order NLS equation~\cite{WLPRE},  fifth-order NLS equation~\cite{AK2},  but also in the coupled systems such as the NLS-MB system~\cite{LiuPLA}, HMB system~\cite{WLJP} and AB system~\cite{WLAB}. The  difference between  {Fig.~2(a)} and {Fig.~2(b)} is that the former shows the single main peak  while the  latter displays the double main  peaks.  In order to more clearly reveal the regularity of transformation between these two waves, we consider $|q(0, 0)|_{zz}^{2}$ as a control variable.  The maximum amplitude of $|q(z, t)|^{2}$ at ($0,0$) can be presented analytically [in Fig.~2(a)],
\begin{equation}\label{AP11}
|q(0,0)|^{2}=\exp(\Gamma(z))(2\,\beta+1)^{2}\,,
\end{equation}
which is related to the  imaginary part of eigenvalue $\beta$ and gain/loss $\Gamma(z)$.
Unfortunately, due to the complexity of $|q(0,0)|_{zz}^{2}$, it's difficult to give it's expression analytically. Thus, we only demonstrate the effect of $\beta$ on $|q(0,0)|_{zz}^{2}$ numerically. As shown in \textbf{Fig.~3}, the green line ($-1.28<\beta<0$) corresponds to the case $|q(0,0)|_{zz}^{2}>0$ that means the ordinate origin ($0, 0$) is a minimum and the soliton has two main peaks with identical amplitude. Nevertheless, if the value of $\beta$ exceeds a certain range, i.e., $\beta>0$ or $\beta<-1.28$, the value of $|q(0,0)|_{zz}^{2}$ is less than zero, which results in the formation of one main peak. Consequently, the transition between these two kinds of waves can be governed by the imaginary part of eigenvalue $\beta$.

\begin{figure}[H]\centering
\subfigure[]{\includegraphics[width=150 bp]{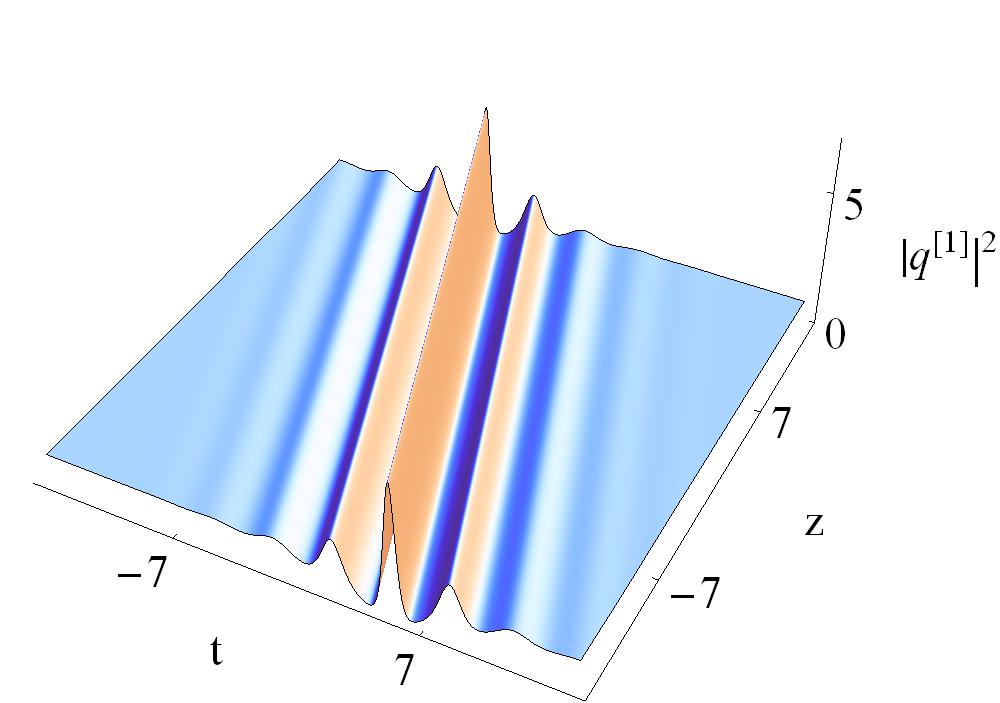}}
\quad
\subfigure[]{\includegraphics[width=150 bp]{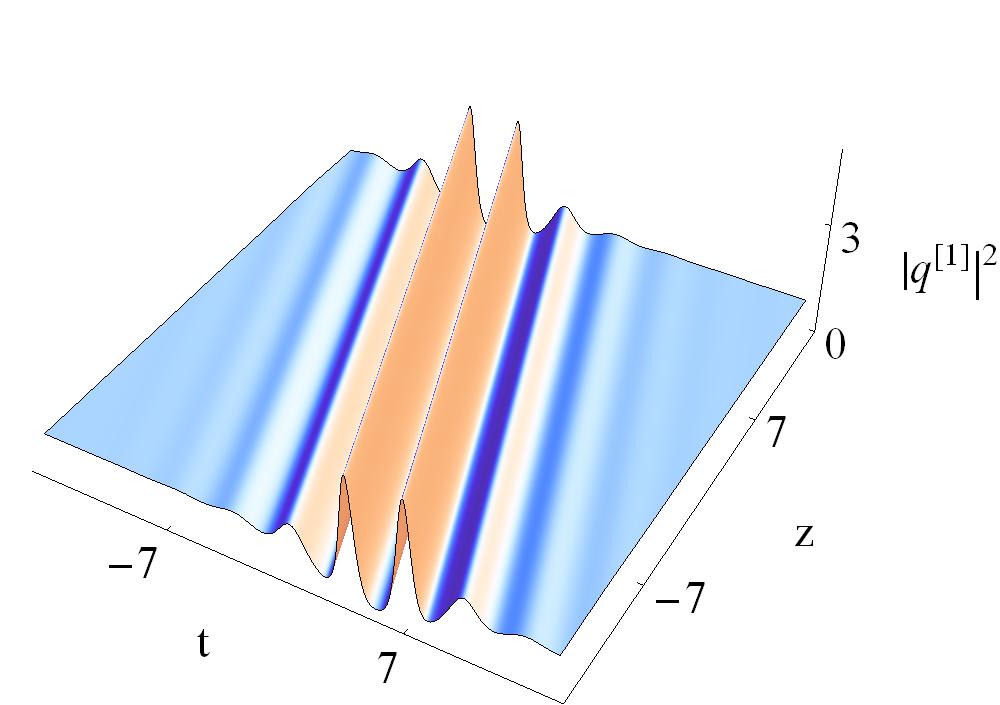}}
\\
\subfigure[]{\includegraphics[width=150 bp]{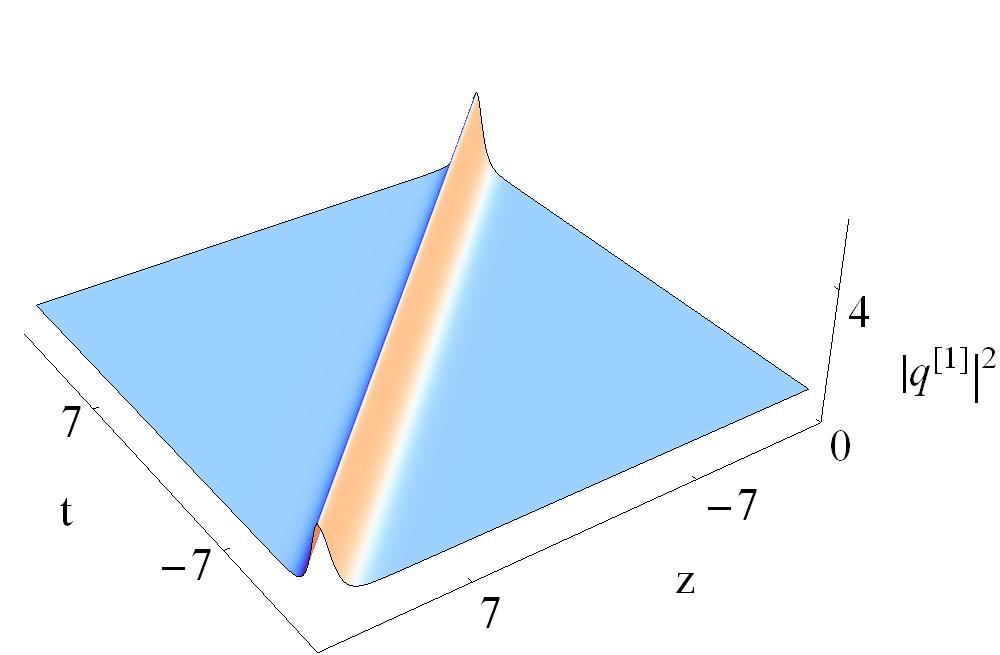}}
\quad
\subfigure[]{\includegraphics[width=150 bp]{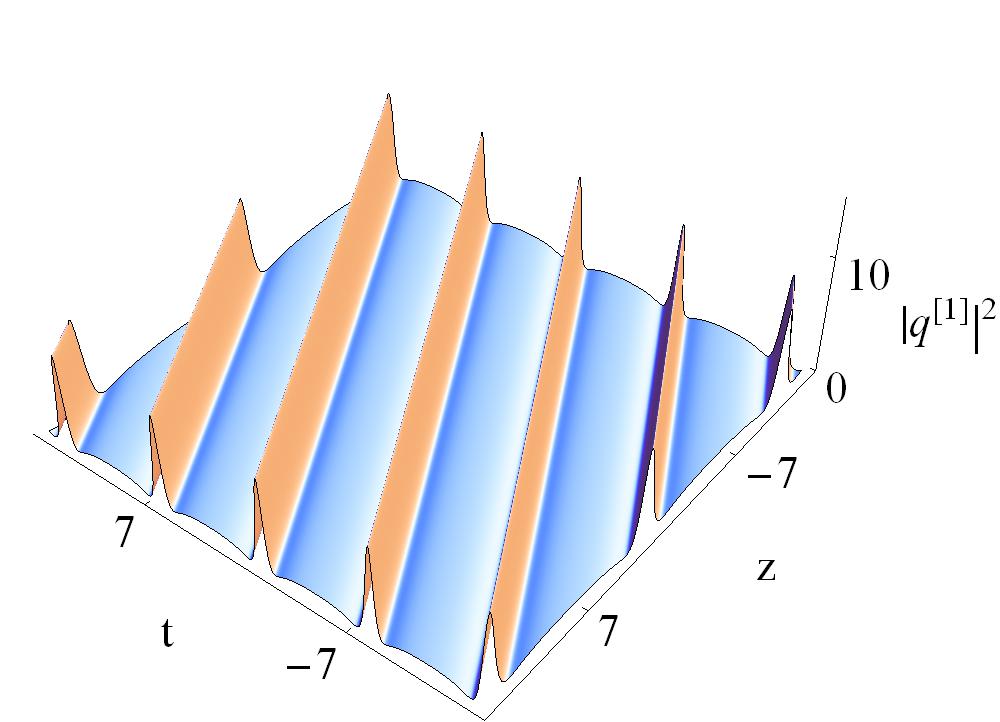}}
\quad
\subfigure[]{\includegraphics[width=150 bp]{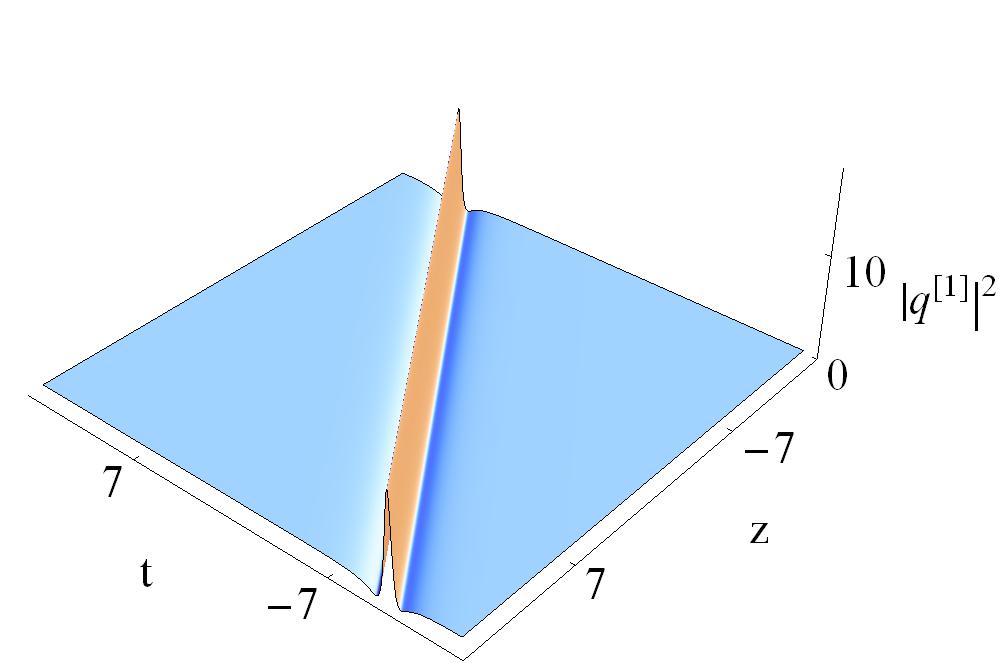}}
\caption{\footnotesize Four types of transformed nonlinear waves with
\,$R(z)=d_{2}(z)=d_{3}(z)=0.1$, (a) Multi-peak soliton (single main peak) with $k_{1}=1,\,k_{2}=-1,\, \lambda_{1}=\lambda_{2}^{*}=0.2+0.6\,i$, (b) M-shaped soliton (double main peak) with $k_{1}=1,\,k_{2}=-1,\, \lambda_{1}=\lambda_{2}^{*}=0.2-0.6\,i$, (c) Antidark soliton with $k_{1}=k_{2}=1,\, \lambda_{1}=\lambda_{2}^{*}=\frac{1}{12}+1.4\,i$, (d) Periodic wave with $k_{1}=1,\,k_{2}=-1,\, \lambda_{1}=\lambda_{2}^{*}=\frac{1}{12}+0.9\,i$, (e) W-shaped soliton with $k_{1}=1,\,k_{2}=-1,\, \lambda_{1}=\lambda_{2}^{*}=\frac{1}{12}+\,i.$}
\end{figure}
\begin{figure}[H]\centering
\quad
\includegraphics[width=180 bp]{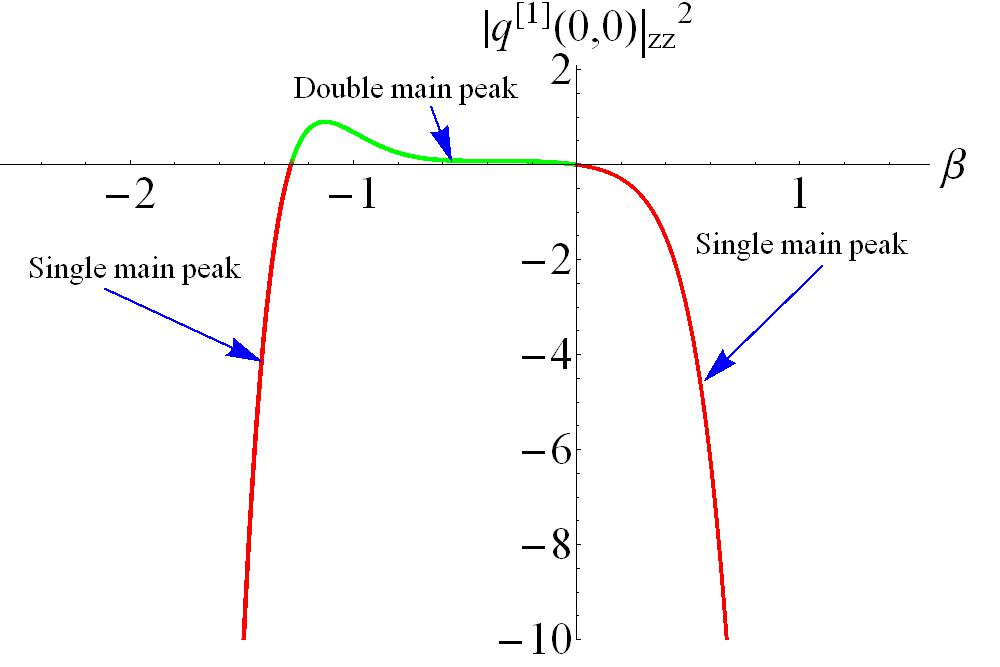}
\caption{\footnotesize Effects of the imaginary part of eigenvalue $\beta$ on $|q(0, 0)|^{2}_{zz}$  with
\,$\delta_{1}=\delta_{2}=\delta_{3}=0.1,\,k_{1}=1,\,k_{2}=-1\,$ and $\,\alpha=0.2$. Two zeros of the $|q(0, 0)|^{2}_{zz}$ are in $(0, 0)$ and $(-1.28, 0)$ respectively. The green line corresponds to the double main peaks while the red line corresponds to the single peaks.}
\end{figure}
Next, we display two special nonlinear wave structures from the solution~(\ref{EXP-B1}), i.e., the antidark soliton
and periodic wave. The former exists in isolation when $h_{R}$ vanishes, while the latter independently exists when $h_{I}$ vanishes. Therefore, the antidark soliton and periodic wave are shown in forms of exponential and trigonometric functions respectively. Specifically, the analytical expressions read as, for the soliton,
\begin{equation}\label{EXP-B1S}
\begin{aligned}
q_{S}^{[1]}=c(z)\,\bigg(1+2\,\beta\,\frac{G_{S}^{[1]}+i\,H_{S}^{[1]}}{D_{S}^{[1]}}\bigg)\,e^{i\,\rho}\,,
\end{aligned}
\end{equation}
 with
 \begin{equation}
\begin{aligned}
&G_{S}^{[1]}=k_{1}k_{2}\,\cosh(2\,\chi_{I})-\cosh(z\,V_{T}+t\,h_{I})\,\sin(2\,\chi_{R})\,,\nonumber \\
& H_{S}^{[1]}=\cos(2\,\chi_{R})\,\sinh(z\,V_{T}+t\,h_{I})\,,\nonumber\\
& D_{S}^{[1]}=\cosh(z\,V_{T}+t\,h_{I})\,\cosh(2\,\chi_{I})-k_{1}k_{2}\,\sin(2\,\chi_{R})\,,\nonumber
\end{aligned}
 \end{equation}
and for the periodic wave,
\begin{equation}\label{EXP-B1P}
\begin{aligned}
q_{P}^{[1]}=c(z)\,\bigg(1+2\,\beta\,\frac{G_{P}^{[1]}+i\,H_{P}^{[1]}}{D_{P}^{[1]}}\bigg)\,e^{i\,\rho}\,,
\end{aligned}
\end{equation}
 with
 \begin{equation}
\begin{aligned}
&G_{P}^{[1]}=k_{1}k_{2}\cos(z\,V_{H}+t\,h_{R})\,\cosh(2\,\chi_{I})-\sin(2\,\chi_{R})\,,\nonumber \\
& H_{P}^{[1]}=k_{1}k_{2}\sin(z\,V_{H}+t\,h_{R})\,\sinh(2\,\chi_{I})\,,\nonumber\\
& D_{P}^{[1]}=\cosh(2\,\chi_{I})-k_{1}k_{2}\cos(z\,V_{H}+t\,h_{R})\,\sin(2\,\chi_{R})\,.\nonumber
\end{aligned}
 \end{equation}
The antidark soliton was firstly reported in the scalar NLS system with the third-order dispersion~\cite{ANS}. Recently, the similar structures have also been found in such coupled models as NLS-MB system~\cite{LiuPLA}, HMB system~\cite{WLJP} and AB system~\cite{WLAB}. The soliton depicted in Fig.~2(c) lies on a plane-wave background with a peak $c(z)^{2}(1+2\,\beta)^{2} $ and will become a standard bright soliton as $c(z)\rightarrow0$.
Fig.~2(d) shows the periodic wave with the period $P=\frac{\pi}{h_{R}}$. Interestingly, in spite of the same expression, the wave in  Fig.~2(d) looks like a higher-order wave.
When the period $h_{R}$ is close to zero, namely, $\beta\rightarrow1$, the periodic wave will reduce to a W-shaped soliton, as shown in Fig.~2(e). In this case, the solution~(\ref{EXP-B1P}) is transformed
into
 \begin{equation}
\begin{aligned}\label{EXP-B1RW}
&q_{RW}^{[1]}=\frac{144\,d_{3}(z)^{2}}{36\,d_{3}(z)^{2}+(12\,d_{3}(z)\,t+z\, d_{2}(z)^{2}+72\,z\,d_{3}(z)^{2})^{2}}+\exp(\frac{1}{2}\Gamma (z))-2\,,
\end{aligned}
 \end{equation}
which is referred to as the W-shaped soliton or a long-lived rogue wave.

We further derive the second-order transformed nonlinear waves. By means of the formulas~(\ref{ndt}) with $N=2$, the two-breather solution of Eq.~(\ref{vcHirota})  is given by
\begin{equation}\label{two-oreder}
\begin{aligned}
 q^{[2]}_{B}=q^{[0]}-2\,i\,\sqrt{\frac{d_{2}(z)}{R(z)}}\,\frac{\Delta_{1}^{[2]}}{\Delta^{[2]}}\,, \end{aligned}
\end{equation}
with
 {\begin{subequations}\begin{gather}
\quad q^{[0]}=c(z)\,e^{i\,\rho}\,,\nonumber\\
\quad \lambda_{1}=\lambda_{2}^{*}=\alpha_{1}+\beta_{1}\,i,\quad
\lambda_{3}=\lambda_{4}^{*}=\alpha_{2}+\beta_{2}\,i, \nonumber\\
\quad \psi_{2}=-\varphi_{1}^{*},\quad
\varphi_{2}=\psi_{1}^{*};\quad
\psi_{4}=-\varphi_{3}^{*},\quad
\varphi_{4}=\psi_{3}^{*};\nonumber \\
\quad
\varphi_{j}={k_{1}}\frac{-i\,h_{j}+2 \,i\, \lambda_{j}+i\,n}{2}\,e^{i\,(\varpi_{j}+\frac{\rho}{2})}+k_{2}\,e^{-i(\varpi_{j}-
\frac{\rho}{2})}\,,\nonumber\\
\quad \psi_{j}=k_{1}\,e^{i\,(\varpi_{j}-\frac{\rho}{2})}+k_{2}\,\frac{-i\,h_{j}+2 \, i \,\lambda_{j}+i \,n }{2}\,e^{-i(\varpi_{j}+\frac{\rho}{2})}\,,\nonumber\\ \quad
   j=1,3,
\quad k_{1}=1,\,k_{2}=\pm1, \nonumber \\
\quad \Delta_{1}^{[2]}=
\begin{vmatrix}{}
\lambda_{1}\varphi_{1}&\varphi_{1}&\lambda_{1}^{2}\varphi_{1}&\psi_{1} \nonumber\\
-\lambda_{2}\varphi_{2}&-\varphi_{2}&-\lambda_{2}^{2}\varphi_{2}&\psi_{2}\nonumber \\
\lambda_{3}\varphi_{3}&\varphi_{3}&\lambda_{3}^{2}\varphi_{3}&\psi_{3}\nonumber \\
-\lambda_{4}\varphi_{4}&-\varphi_{4}&-\lambda_{4}^{2}\varphi_{4}&\psi_{4}
\end{vmatrix}\,,
\quad\quad
\Delta^{[2]}=
\begin{vmatrix}{}
\lambda_{1}\varphi_{1}&\varphi_{1}&\lambda_{1}\psi_{1}&\psi_{1} \nonumber\\
-\lambda_{2}\varphi_{2}&-\varphi_{2}&\lambda_{2}\psi_{2}&\psi_{2}\nonumber \\
\lambda_{3}\varphi_{3}&\varphi_{3}&\lambda_{3}\psi_{3}&\psi_{3}\nonumber \\
-\lambda_{4}\varphi_{4}&-\varphi_{4}&\lambda_{4}\psi_{4}&\psi_{4}
\end{vmatrix}\,.
\end{gather}\end{subequations}}
By using the solution~(\ref{two-oreder}) and transition condition~(\ref{KZFC}), we can obtain various nonlinear interactions among different types of transformed nonlinear waves. For more detailed analysis of these nonlinear interactions, one can  refer to Refs.~\cite{WLPRE,WLJP,LiuAOP}. Hereby, we only exhibit a few typical examples. \textbf{Figs.~4(a)$\sim$4(c)} describe the interactions between  two multi-peak solitons, and between the periodic wave and antidark soliton.
We note that the collisions are elastic, i.e., the waves restore the original shapes, amplitudes and velocities after each collision with a small phase shift. On the other hand,
employing the semirational forms of the solution~(\ref{two-oreder}) and transition condition~(\ref{KZFC}), we have the interactions between the W-shpaed  soliton and the antidark soliton, and between two W-shaped soitons, which are depicted in\textbf{ Fig.~5}.

\begin{figure}[H]\centering
\subfigure[]{\includegraphics[width=150 bp]{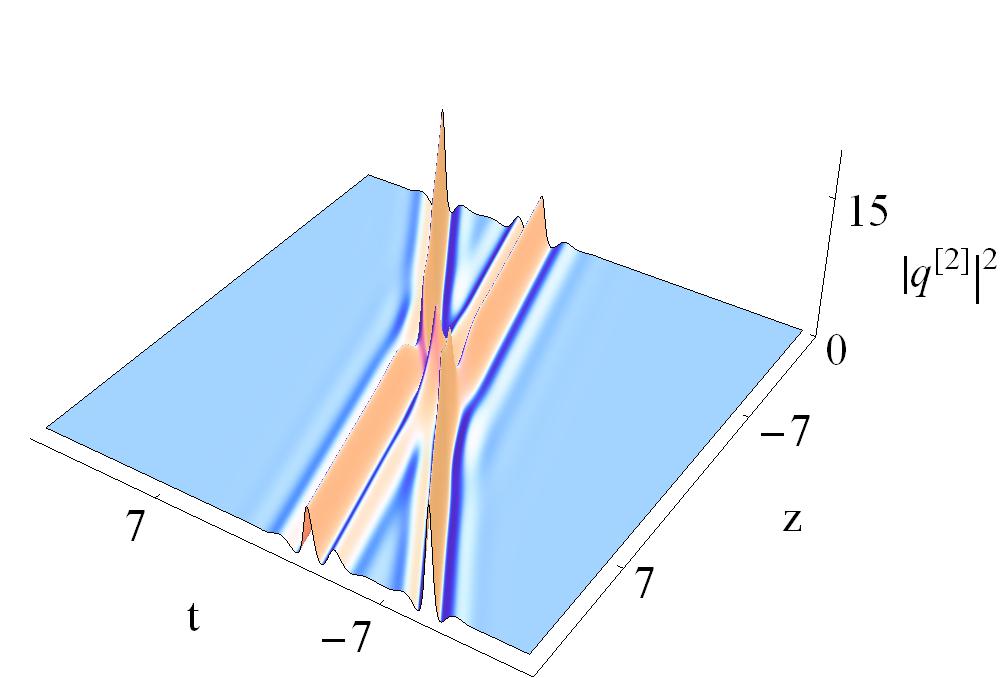}}
\quad
\subfigure[]{\includegraphics[width=150 bp]{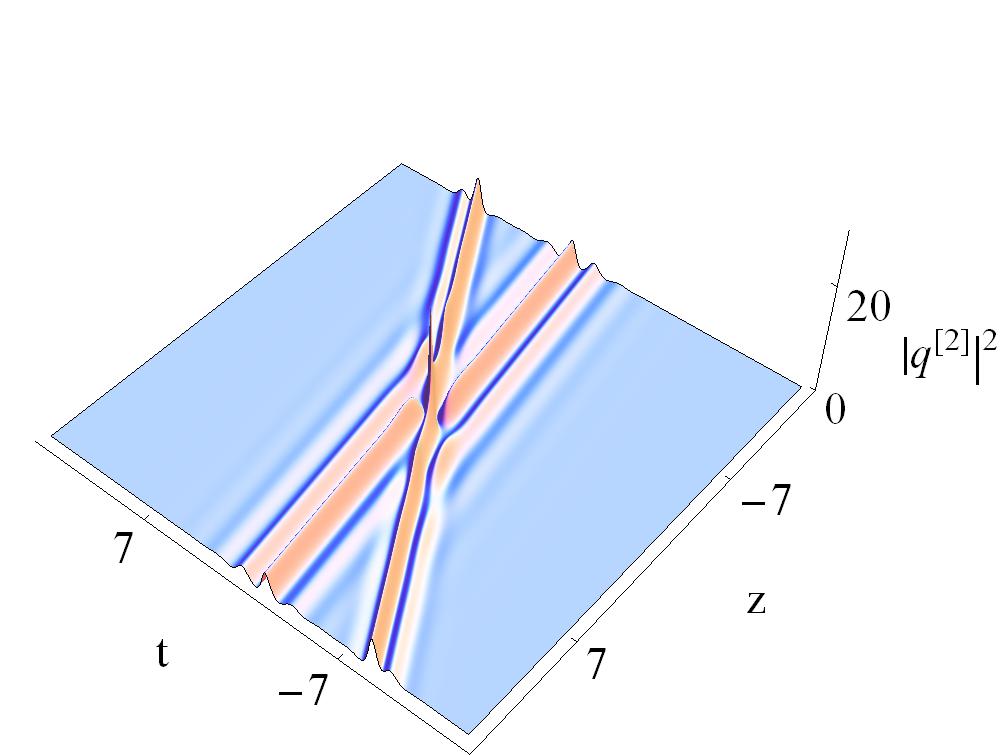}}
\quad
\subfigure[]{\includegraphics[width=150 bp]{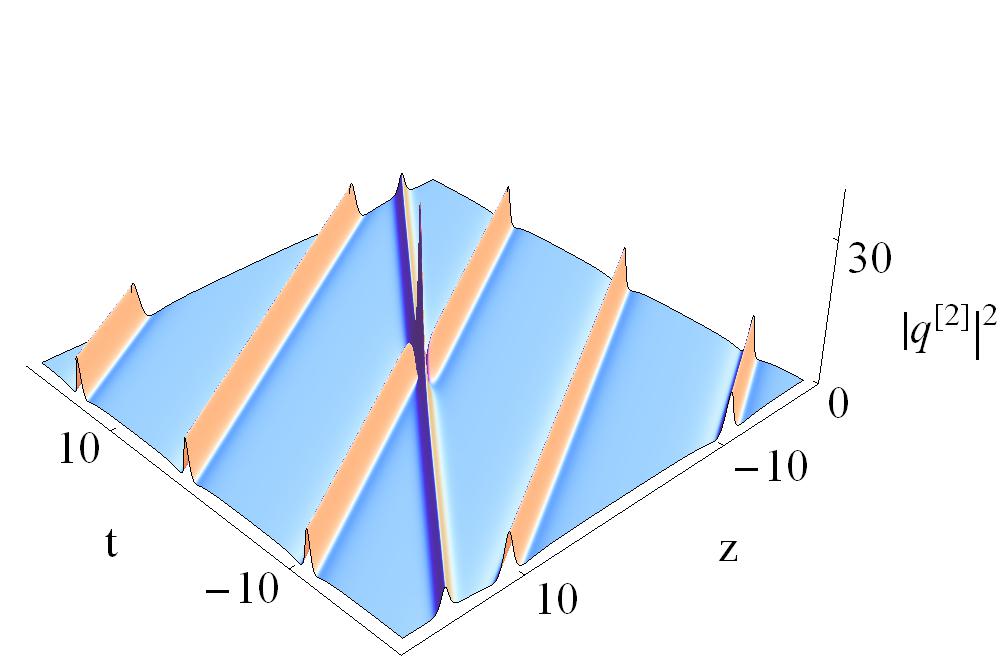}}
\caption{\footnotesize The interactions between two nonlinear waves with
\,$\delta_{1}=\delta_{2}=\delta_{3}=0.1$,\, (a) The collision between two multi-peak soliton with $k_{1}=k_{2}=1,\, \lambda_{1}=\lambda_{2}^{*}=0.5+1.2\,i,\, \lambda_{3}=\lambda_{4}^{*}=0.5+0.8\,i$, (b) The collision between two M-shaped soliton with $k_{1}=1,\,k_{2}=-1,\, \lambda_{1}=\lambda_{2}^{*}=0.5+1.2\,i,\,\lambda_{3}=\lambda_{4}^{*}=0.5+0.6\,i,$ (c) The collision between antidark soliton and periodic wave with $k_{1}=1,\,k_{2}=-1,\, \lambda_{1}=\lambda_{2}^{*}=\frac{1}{12}+1.6\,i,\,\lambda_{3}=\lambda_{4}^{*}=\frac{1}{12}+0.97\,i.$}
\end{figure}

\begin{figure}[H]\centering
\subfigure[]{\includegraphics[width=150 bp]{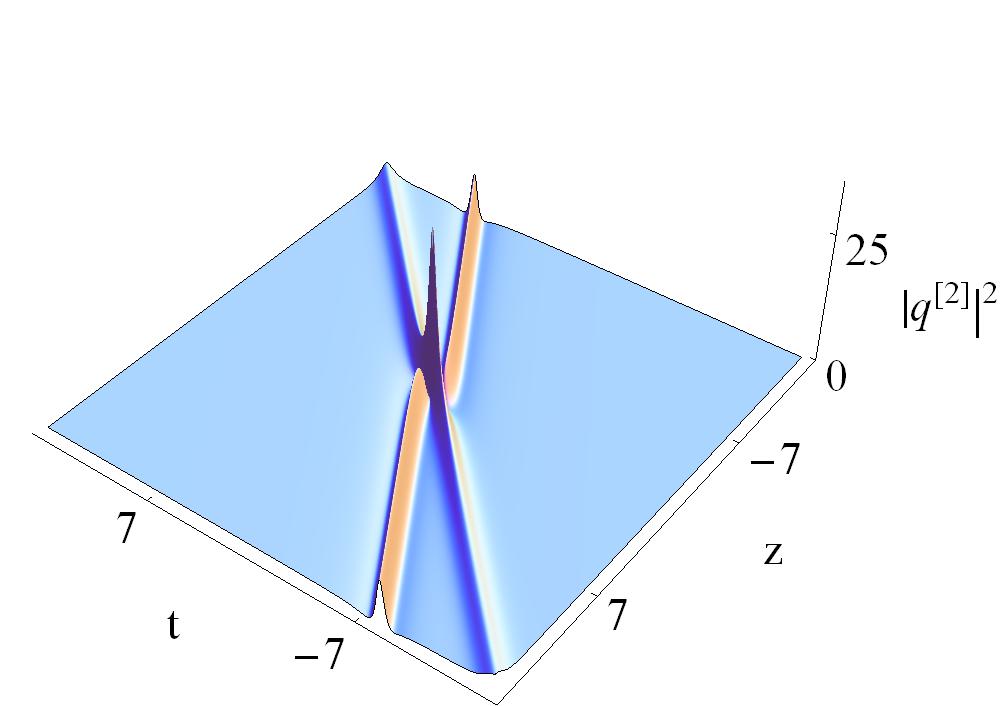}}
\quad
\subfigure[]{\includegraphics[width=150 bp]{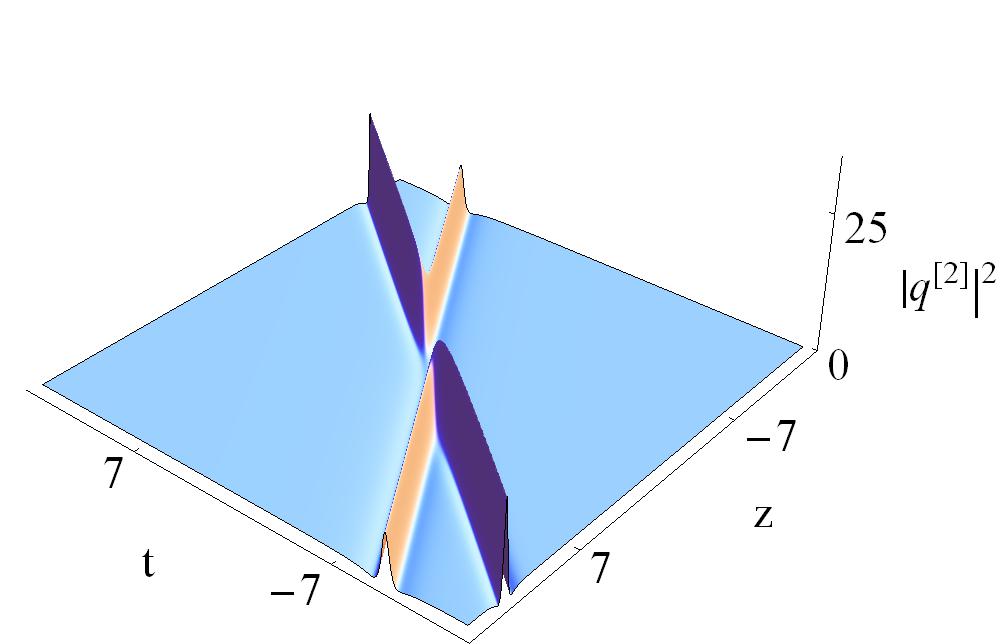}}
\caption{\footnotesize The interactions between two nonlinear waves with
\,$\delta_{1}=\delta_{2}=\delta_{3}=0.1,\,\,$(a) The collision between W-shaped soliton and antidark soliton with $k_{1}=k_{2}=1,\, \lambda_{1}=\lambda_{2}^{*}=\frac{1}{12}+\,i,\, \lambda_{3}=\lambda_{4}^{*}=\frac{1}{12}-1.6\,i,$ (b) The collision between two W-shaped soliton with $k_{1}=k_{2}=1,\, \lambda_{1}=\lambda_{2}^{*}=\frac{1}{12}+1.2\,i,\, \lambda_{3}=\lambda_{4}^{*}=\frac{1}{12}+1.4\,i.$}
\end{figure}

\textbf{B. Dispersion management  and nonlinear management}

We discuss the effects of the variable coefficients on the nonlinear waves, i.e., the dispersion management and nonlinear management. Here we will pay our attention to the multi-peak soliton with single main peak. We omit the results on other types of nonlinear waves, since the effects of the variable coefficients on them
are similar to the multi-peak soliton  with the same initial physical parameters.
From the constraints~(\ref{cc}), we note that five variable coefficients $d_{2}(z)$, $d_{3}(z)$, $R(z)$, $\gamma(z)$ and $\Gamma(z)$ are not independent of each other.
The  ratio of $d_{2}(z)$ and $R(z)$ must be equal to that of $d_{3}(z)$ and $\gamma(z)$. Additionally, $\Gamma(z)$ should be expressed by $d_{2}(z)$ and $R(z)$ (or $d_{3}(z)$ and $\gamma(z)$). Therefore, we will consider dispersion management  and nonlinear management of the multi-peak soliton under the integrability condition~(\ref{cc}).

We first study the effect of TOD coefficient $d_{3}(z)$ and time-delay correlation to the cubic term $\gamma(z)$ on the multi-peak soliton. We fix the values of $d_{2}(z)$ and $R(z)$ while change the values of $d_{3}(z)$  and $\gamma(z)$. For simplicity, we set $\Gamma(z)=0$. As shown in \textbf{Fig.~6}, we observe a  compressed effect of the multi-peak soliton by increasing the values of  $d_{3}(z)$ and $\gamma(z)$. In other words, a further increase of the values of  $d_{3}(z)$ and $\gamma(z)$ leads to stronger localization and a smaller oscillation period. However, the amplitudes of the multi-peak soliton including the maximum and minimum ones do not change markedly, especially for the main peak. It's worth pointing out that we have to increase or decrease the values of $d_{3}(z)$ and $\gamma(z)$ simultaneously to  ensure the integrability condition~(\ref{cc})  to be true.

\begin{figure}[H]\centering
{\includegraphics[width=190 bp]{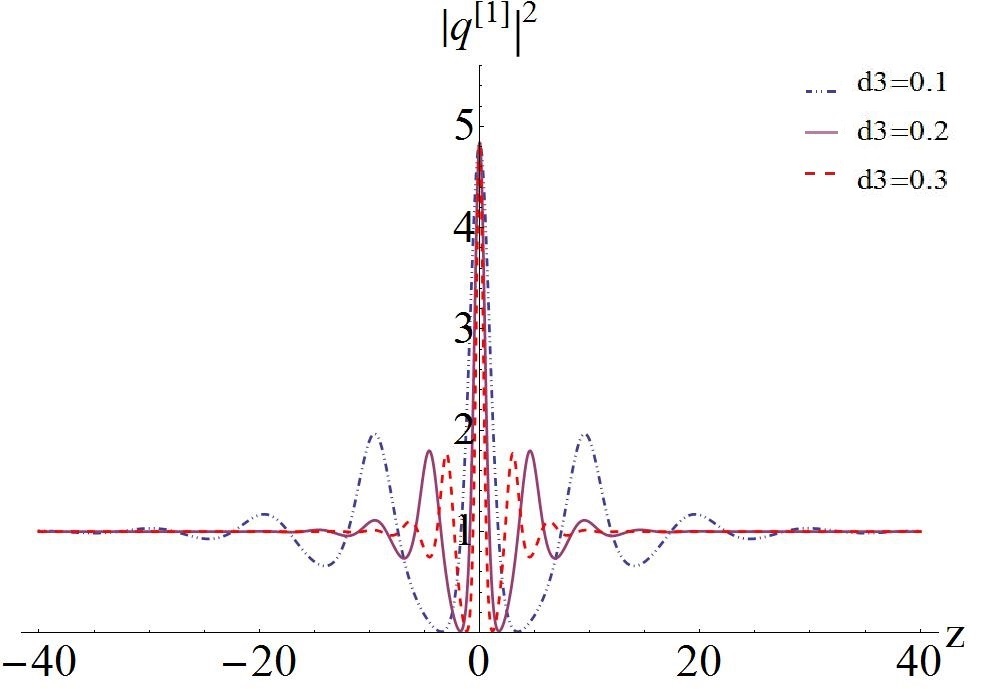}}
\caption{\footnotesize The compression effect of TOD coefficient on the multi-peak solitons with
\,$\delta_{1}=\delta_{2}=0.1,\,k_{1}=1,\,k_{2}=-1,\, \lambda_{1}=\lambda_{2}^{*}=0.2+0.6\,i.$}
\end{figure}

Secondly, we investigate how  the GVD coefficient $d_{2}(z)$ and Kerr nonlinear coefficient $R(z)$ influence on the multi-peak soliton. In this case, the values of $d_{3}(z)$  and $\gamma(z)$ cannot be changed and $d_{2}(z)$ and $R(z)$ are various. From \textbf{Fig.~7}, we discover that increasing the values of $d_{2}(z)$ and $R(z)$ also results in  stronger localization and a smaller oscillation period for the multi-peak soliton. More interestingly, different from the TOD effects, the GVD coefficient can affect the peak number of the multi-peak soliton.
When $d_{2}(z)=R(z)=1$, the  soliton has seven humps  (see the dotted line in Fig.~7). By raising the values of $d_{2}(z)$ and $R(z)$, we can observe that the humps of the soliton  increase from seven to thirteen (see the solid line in Fig.~7).
Further, by compare with the main peak, the amplitudes of secondary ones increase obviously.
\begin{figure}[H]\centering
\includegraphics[width=190 bp]{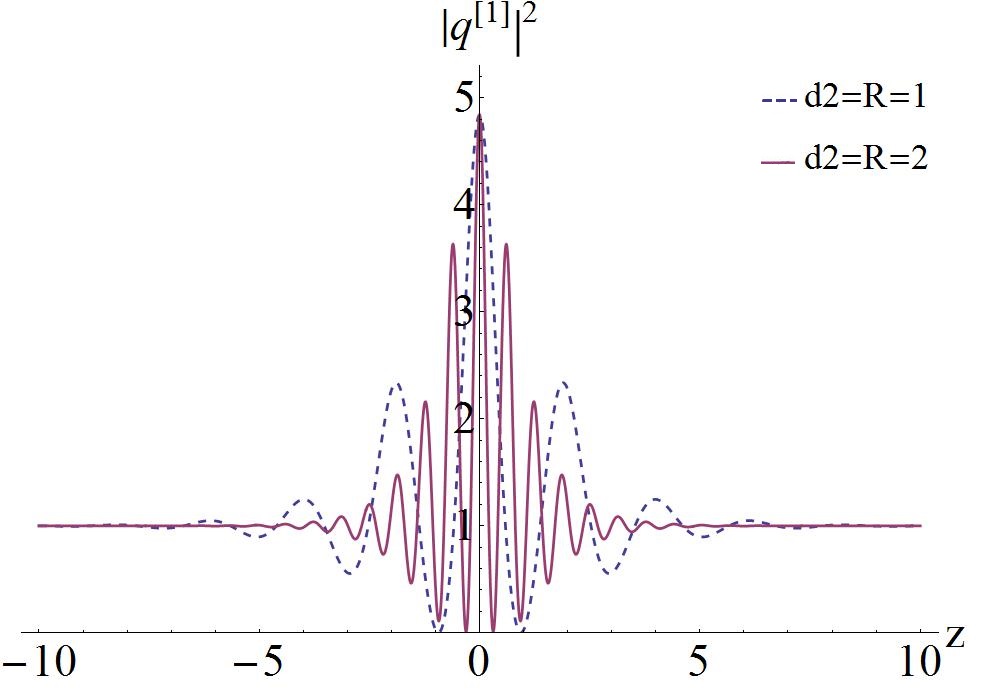}
\caption{\footnotesize  The effects of GVD coefficient on the peak number of multi-peak solitons with
\,$\delta_{3}=0.1,\,k_{1}=1,\,k_{2}=-1,\, \lambda_{1}=\lambda_{2}^{*}=0.2+0.6\,i.$}
\end{figure}

Thirdly, we discuss the effects of the  gain/loss coefficient $\Gamma(z)$. It should be pointed out that the GVD coefficient $d_{2}(z)$ and Kerr nonlinear coefficient $R(z)$
need to meet non-linear relation ($d_{2}(z)=k(z)\,R(z)$, $k(z)$ is a function of $z$) because the case $d_{2}(z)=k \,R(z)$ will lead to the vanishing gain/loss effect. Based on the fact that decreasing GVD in a fiber has been realized, as an example, we consider an exponential dispersion decreasing fiber system  with
\begin{equation}
d_{2}(z)=\delta_{1}\exp(\xi\,z),\qquad d_{3}(z)=\delta_{3}\exp(\xi\,z), \qquad  R(z)=\delta_{1},\qquad \gamma(z)=\delta_{3},
\end{equation}
\begin{equation}
\Gamma(z)=\xi\,,
\end{equation}
where $\delta_{1}$ ($\delta_{3}$) is the parameter  related to the Kerr nonlinear (TOD) and
$\xi$ denotes the constant net gain or loss.  \textbf{Fig.~8} describes the propagation of a multi-peak soliton whose amplitudes, background and velocity vary due to the  nonvanishing  gain or loss.   From the expression~(\ref{EXP-B1}), we see that the amplitudes of the multi-peak soliton are determined by
$c(z)=\exp(\xi \,t)$,
and the velocity is influenced by
$d_{2}(z)=\delta_{1}\exp(\xi\,z), d_{3}(z)=\delta_{3}\exp(\xi\,z)$.
If $\xi>0$, the amplitude of this wave will increase exponentially whereas it will decrease exponentially. In addition, we observe that the multi-peak soliton is compressed  during the propagation owing to  the exponential dispersion decreasing coefficients. The cases $\xi<0$ and $\xi>0$, respectively,  correspond to the compression and amplification.

Finally, we   consider a soliton management system similar to that of
Ref.~\cite{PP}, i.e., the periodic distributed system
\begin{equation}
R(z)=\delta_{1}\sin(\xi\,z)\,,\quad d_{2}(z)=\delta_{2}\sin(\xi\,z)\,,\quad d_{3}(z)=\delta_{3}\sin(\xi\,z)\,,\quad \gamma(z)=\frac{\delta_{1}\delta_{3}}{\delta_{2}}\sin(\xi\,z)\,.
\end{equation}
 Trigonometric functions are physically relevant because they provide for alternating regions of positive and negative dispersion and nonlinearity, indicated in the improved stability of the solitons~\cite{L1}. The periodically accelerating or decelerating multi-peak soliton  are shown in\textbf{ Fig .~9(a)}. The two solitons in\textbf{ Fig.~9(b) }propagate with periodic oscillation along the time $z$, and the separated solitons collide periodically.

\begin{figure}[H]\centering
\subfigure[]{\includegraphics[width=170 bp]{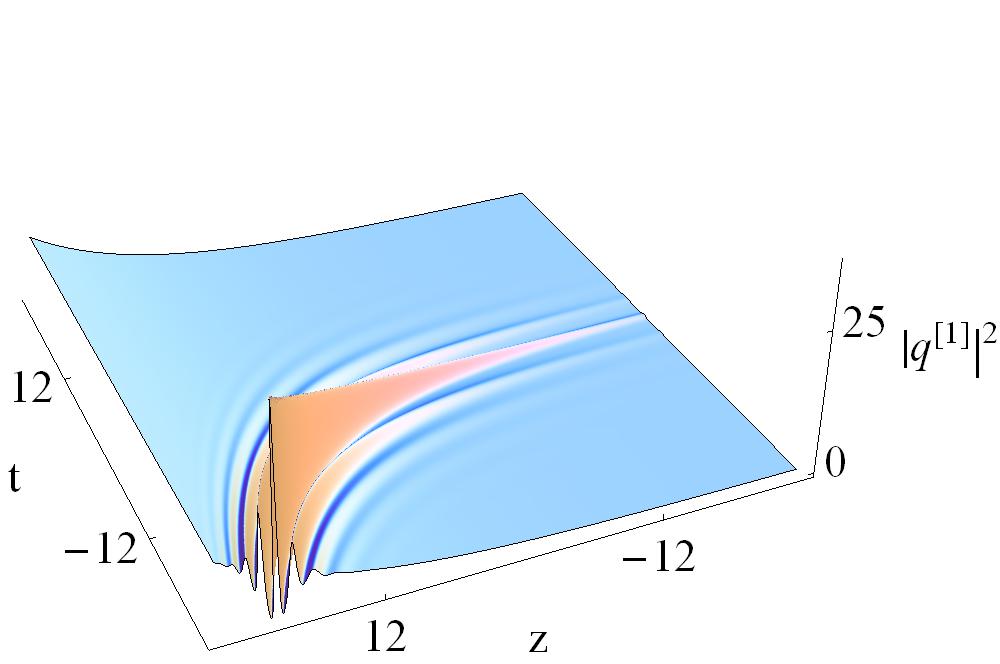}}
\quad
\subfigure[]{\includegraphics[width=100 bp]{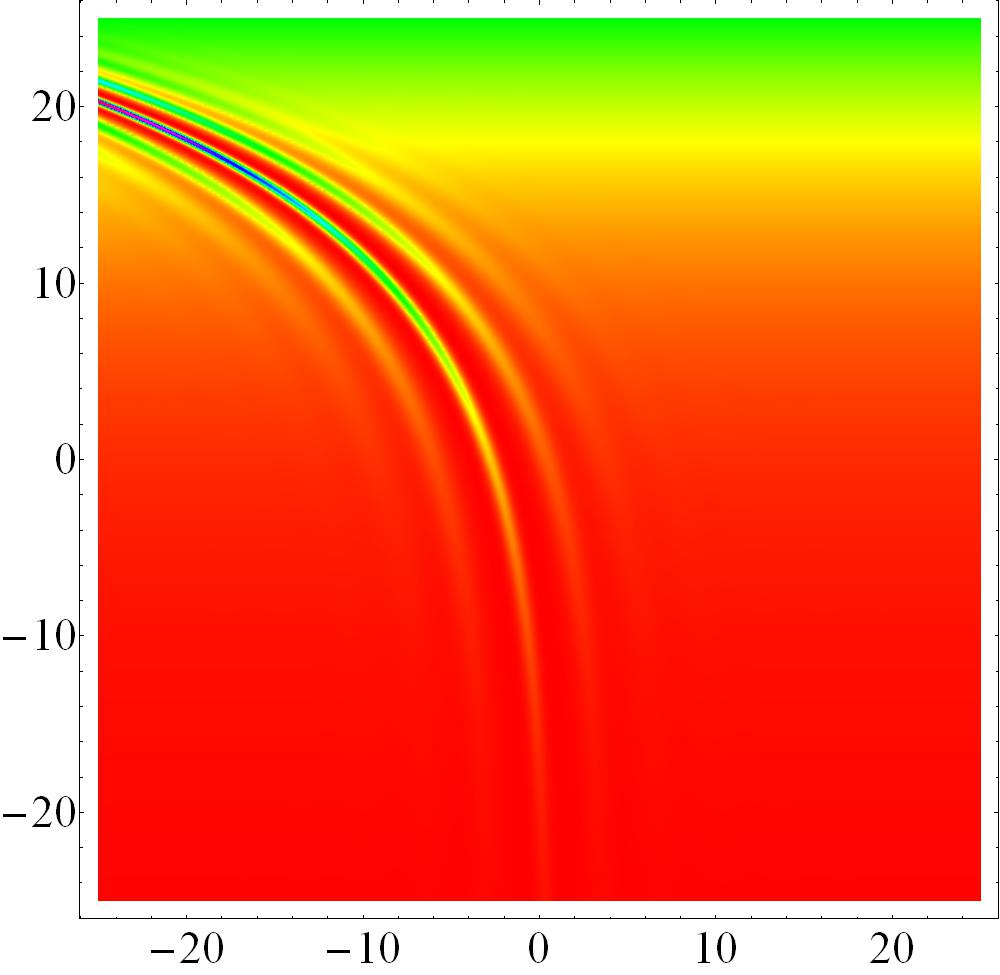}
\includegraphics[height=100 bp]{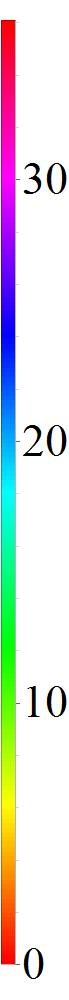}}
\quad
\subfigure[]{\includegraphics[width=170 bp]{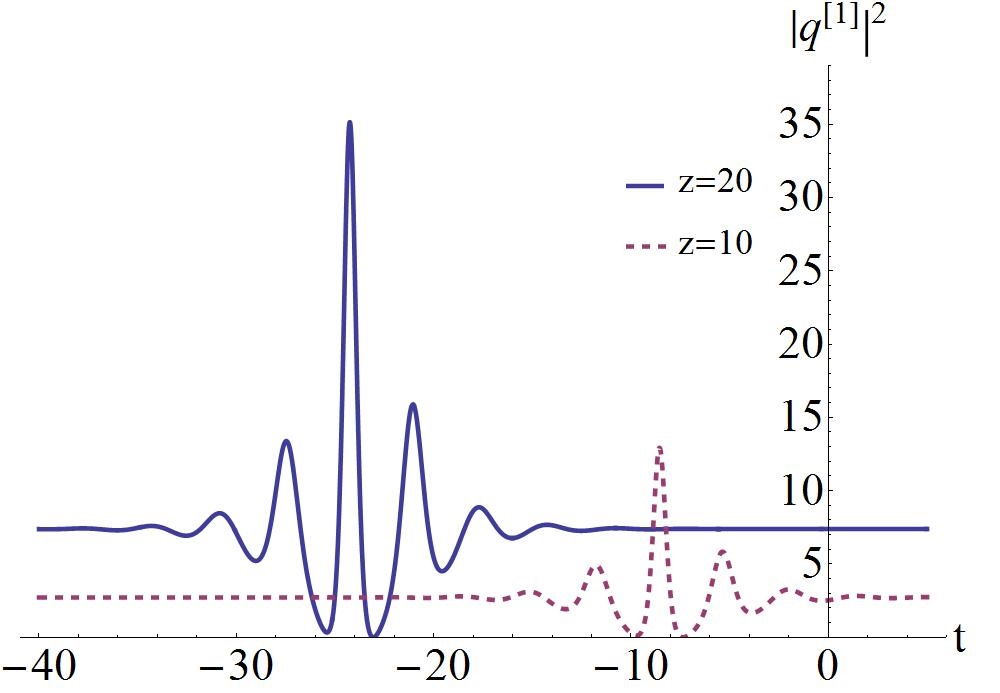}}
\caption{\footnotesize The  effect of gain and loss coefficient on the multi-peak solitons with
\,$\delta_{1}=\delta_{3}=0.1,\,\xi=0.1,\,k_{1}=k_{2}=1\,$ and $\, \lambda_{1}=\lambda_{2}^{*}=0.2+0.6\,i.$ (b) is the density plot of (a). (c) is the cross-sectional view of (a) at $z=10$ and $z=20$.}
\end{figure}

\begin{figure}[H]\centering
\subfigure[]{{\includegraphics[width=150 bp]{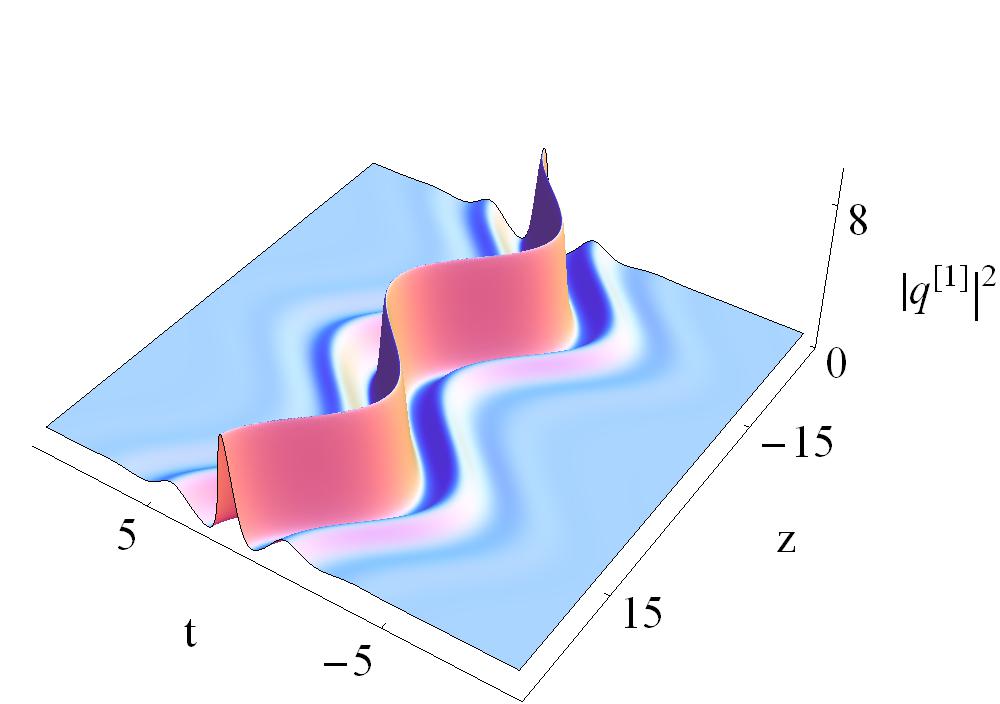}}}
\subfigure[]{{\includegraphics[width=150 bp]{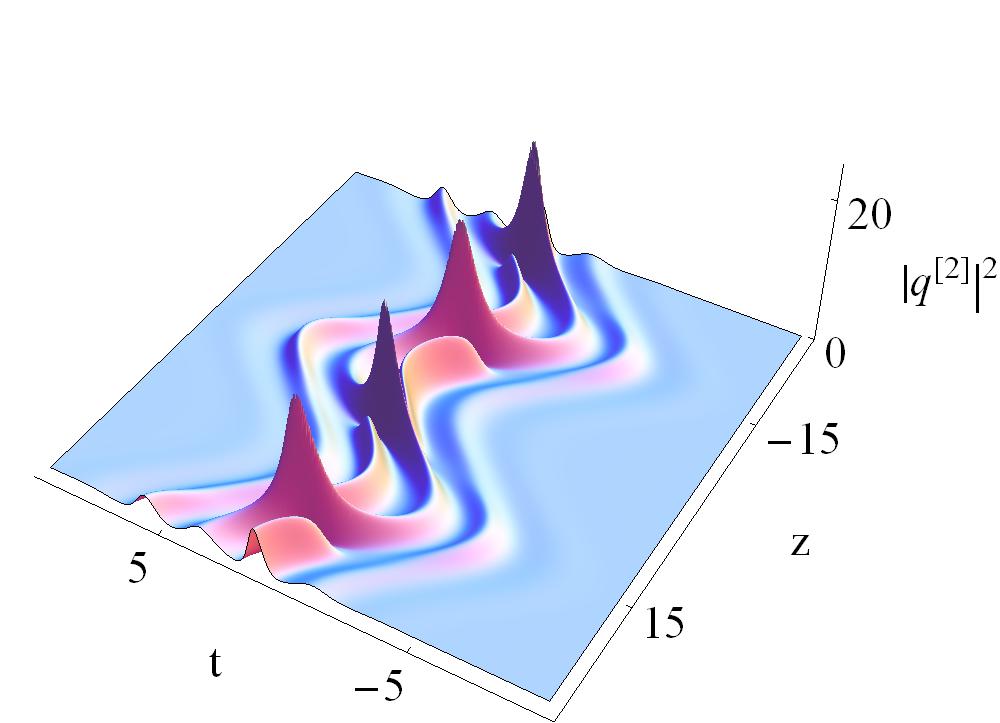}}}
\caption{\footnotesize Periodic variable-motion of the multi-peak solitons with
\,$\delta_{1}=\delta_{2}=\delta_{3}=0.1,\,\xi=0.2,\,$ (a) The first-order multi-peak soliton with $k_{1}=1,\,k_{2}=-1,\, \lambda_{1}=\lambda_{2}^{*}=0.3+0.7\,i,$ (b) The periodic collision between two multi-peak solitons with $k_{1}=1,\,k_{2}=-1,\, \lambda_{1}=\lambda_{2}^{*}=0.3+\,i,\,\lambda_{3}=\lambda_{4}^{*}=0.3+0.8\,i.$}
\end{figure}


\vspace{5mm}
\noindent\textbf{\Large{\uppercase\expandafter{3}. MI characteristics}}

In this section, we reveal the explicit relation between the transition and the distribution characteristics of  MI growth rate for Eq.~(\ref{vcHirota}).  Eq.~(\ref{vcHirota}) admits the following continuous wave solutions,
\begin{equation}\label{EXP-MI1}
\begin{aligned}
\qquad\qquad\quad\quad\quad&{q}(z,t)=c(z)\,e^{i\,(m(z)+n\,t)}=\sqrt{\frac{d_{2}(z)}{R(z)}}\,e^{i\,(m(z)+n\,t)}\,,
\end{aligned}
\end{equation}
where $ n $ are real parameters. A perturbed nonlinear background can be expressed as
\begin{equation}\label{EXP-MI2}
\begin{aligned}
\qquad\qquad\qquad\qquad\quad\quad&{q}(z,t)=(c(z)+\epsilon\,{\widehat{q}}(z,t))\,e^{i\,(m(z)+n\,t)},
\end{aligned}
\end{equation}
into Eq.~(\ref{vcHirota})  yields the evolution equation for the perturbations as
\begin{equation}
\label{lse}
\begin{aligned}
&-i\,R(z)\,\widehat{q}\,d_{2}(z)_{z}+d_{2}(z)^{2}R(z)\Big(2\,\widehat{q}+2\,\widehat{q}^{*}+2\,i\,n\,\widehat{q}_{t}+\widehat{q}_{tt}\Big)+d_{2}(z)\Big(2\,d_{3}(z)R(z)\big(6\,n\,\widehat{q}+6\,n\,\widehat{q}^{*}\\
&-6\,i\,\widehat{q}_{t}+3\,i\,n^{2}\widehat{q}_{t}+3\,n\,\widehat{q}_{tt}-i\,\widehat{q}_{ttt}\big)+i\,\big(\widehat{q}\,R(z)_{z}+2R(z)\widehat{q}_{z}\big)\Big)=0\,.
\end{aligned}
\end{equation}
Noting the linearity of Eq.~(\ref{lse})  with respect to $\widehat{q}$, we introduce
\begin{equation}
\label{lse1}
\begin{aligned}
\qquad\qquad\qquad&{\widehat{q}}(z,t)=u\,c(z)\,e^{i\,(Q\,t-\omega(z))}+v\,c(z)\,e^{-i\,(Q\,t-\omega^{*}(z))}\,,\\
\end{aligned}
\end{equation}
which is characterized by the wave number $\omega$ and frequency $Q$. Using Eq.~(\ref{lse1}) into Eq.~(\ref{lse}) gives a linear homogeneous system of equations for $u$  and $v$:
\begin{equation}\label{18}
\begin{aligned}
\qquad\qquad&d_{2}(z)-n\,Q\,d_{2}(z)-\frac{1}{2}Q^{2}\,d_{2}(z)+6\,n\,d_{3}(z)+6\,Q\,d_{3}(z)\\
&-3\,n^{2}\,Q\,d_{3}(z)-3\,n\,Q^{2}\,d_{3}(z)-Q^{3}\,d_{3}(z)+\omega_{z}(z)=0,
\end{aligned}
\end{equation}
\begin{equation}\label{19}
\begin{aligned}
\qquad\qquad\qquad\qquad d_{2}(z)+6\,n\,d_{3}(z)=0.
\end{aligned}
\end{equation}
From the determinant of the coefficient matrix of Eqs.~(\ref{18})$\sim$(\ref{19}), the dispersion relation for the linearized disturbance can be determined as
{\begin{equation}\begin{aligned}
&\qquad n Q d_2(z) \left(d_3(z) \left(Q^3-6 n^2 Q\right)+2 \omega_{z}(z)\right)+2 Q
   d_3(z) \left(3 n^2+Q^2-6\right) \omega_{z}(z)
   \\&+\frac{1}{4} Q^2 d_2(z){}^2
   \left(-4 n^2+Q^2-4\right)-Q^2 d_3(z){}^2 \left(9 n^4-3 n^2
   Q^2+\left(Q^2-6\right)^2\right)-\omega_{z}(z)^2=0.
\end{aligned}\end{equation}}
Solving the above equation, we have
\begin{equation}
\qquad\quad\omega(z)=Q\,\Big(d_{2}(z)\,n+d_{3}(z)\big(-6+3\,n^{2}+Q^{2}\big)\Big)\pm\frac{1}{2}|Q|\sqrt{(d_{2}(z)+6\,d_{3}(z)\,n)^{2}\,(-4+Q^{2})}.
\end{equation}
In this case, the wave number $\omega(z)$ becomes complex and the disturbance will grow with time exponentially if and only if $Q < Q_{c}=2$, and the growth rate of the instability is given by
\begin{equation}
\qquad\quad\Omega(z)=\frac{1}{2}Q^{2}\sqrt{(d_{2}(z)+6\,d_{3}(z)\,n)^{2}\,(\frac{4}{Q^{2}}-1)}.
\end{equation}
To obtain the maximum growth rate of the instability, we take the derivative of Eq.~(23) with respect to $Q$, and set it to zero. Then, we obtain $Q_{max}=\sqrt{2}$ and the following maximum growth rate of the instability:
\begin{equation}
\qquad\quad\Omega(z)_{max}=\sqrt{(d_{2}(z)+6\,d_{3}(z)\,n)^{2}}.
\end{equation}

\begin{figure}[H]\centering
{\includegraphics[width=150 bp]{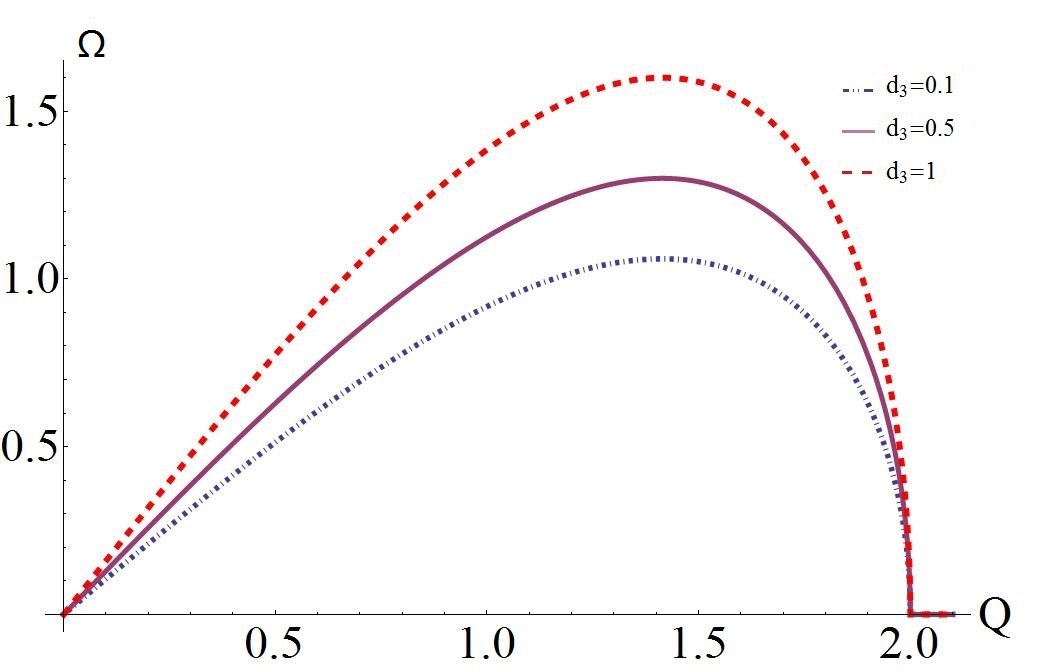}}
\caption{\footnotesize The effects of $d_{3}$ on the  growth rate of  instability with $n=0.1,\,d_{2}=1$. }
\end{figure}

\begin{figure}[H]\centering
 {\subfigure[]{\includegraphics[width=120 bp]{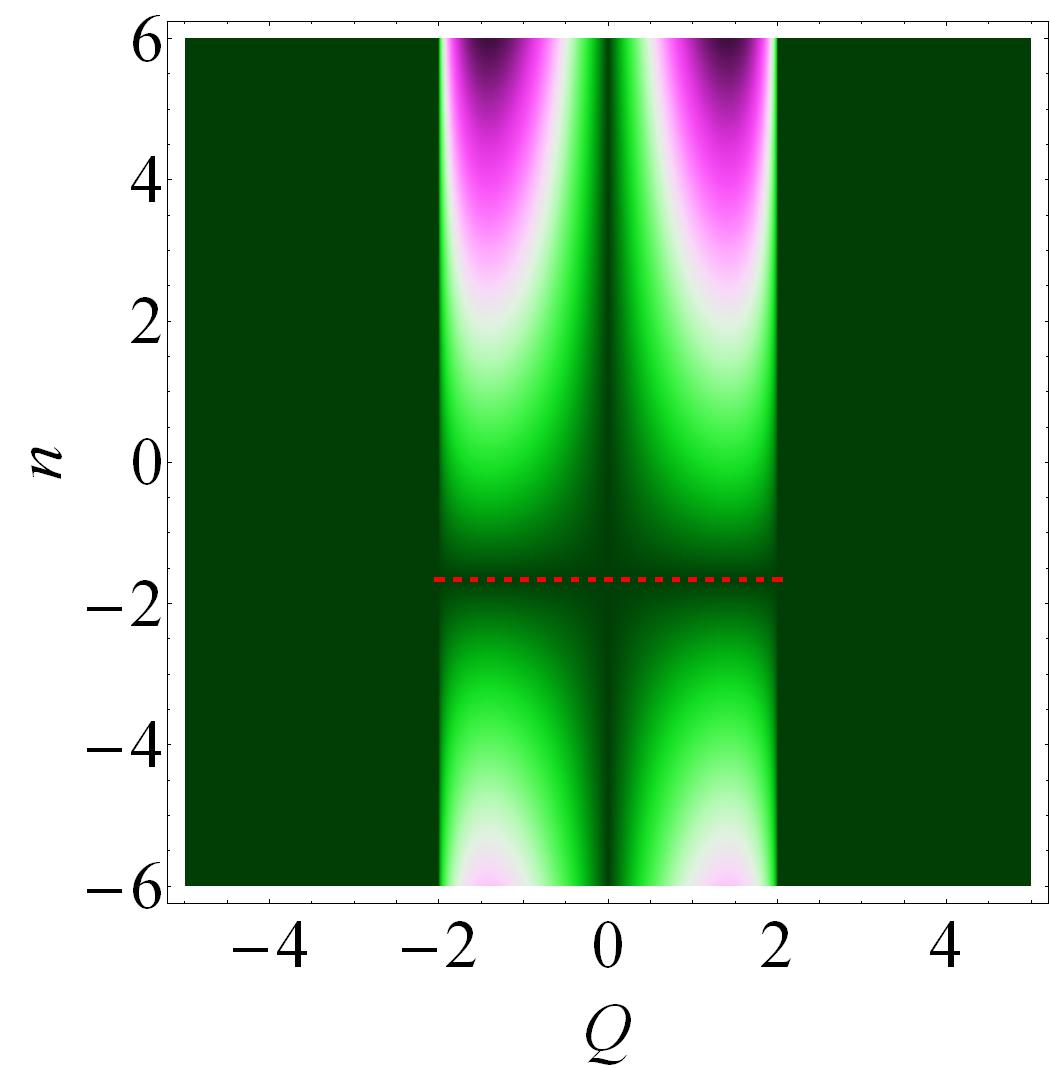}
  \includegraphics[height=120 bp]{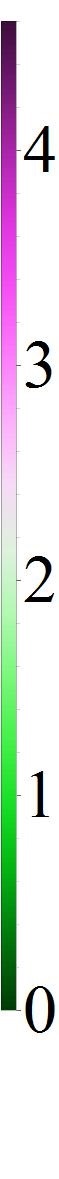}}}\qquad\qquad
 {\subfigure[]{\includegraphics[width=120 bp]{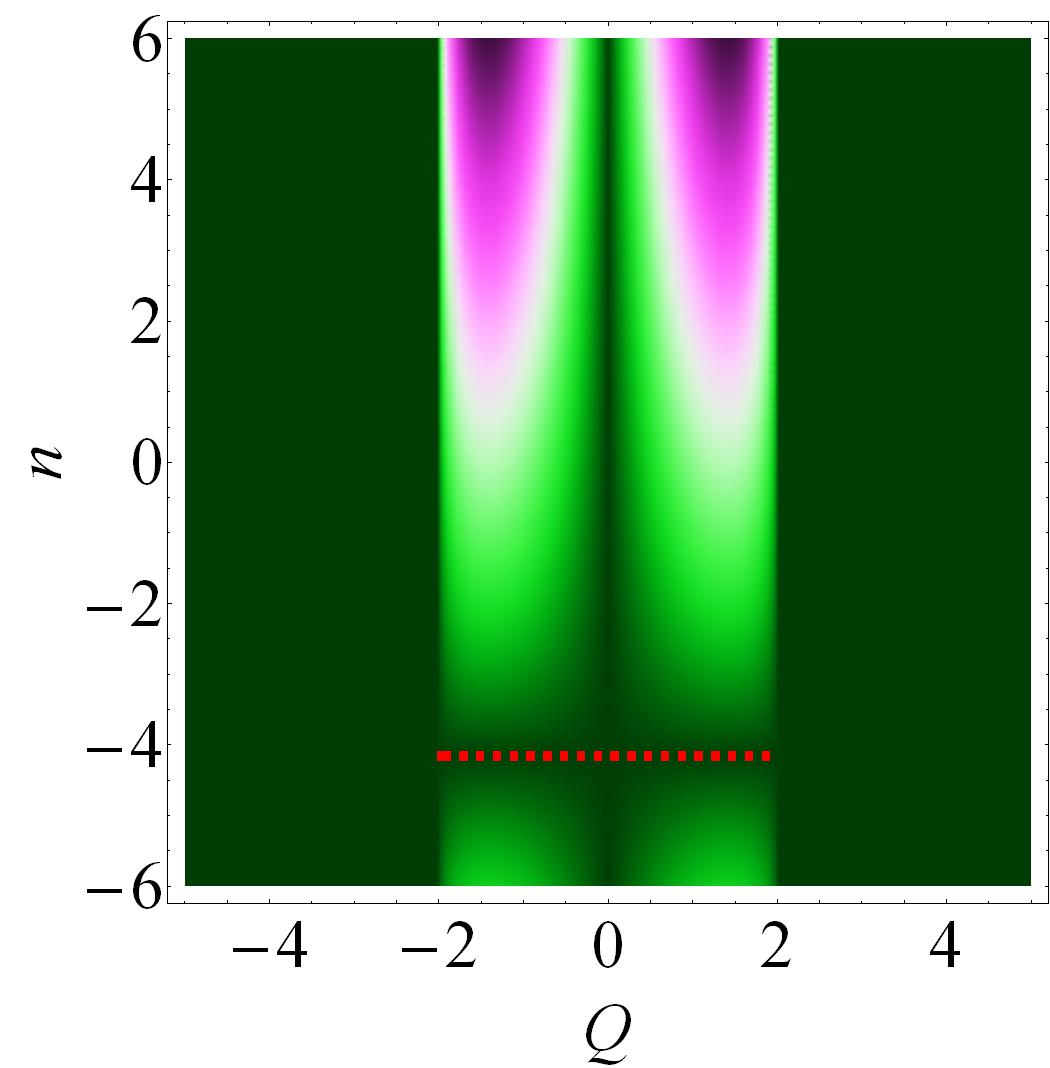}
  \includegraphics[height=120 bp]{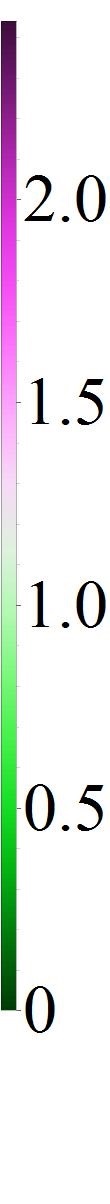}}}
  \caption{\footnotesize Characteristics of MI growth rate $\omega$ on $(Q,n)$ plane with $d_{2}(z)=1$ and $(a)d_{3}(z)=0.1;(b)d_{3}(z)=0.04$. Here the dashed red lines represent  the stability region in
the perturbation frequency region $-2<Q<2 $ , which is given as\, $n=-\frac{d_{2}(z)}{6\,d_{3}(z)}$.}
\end{figure}

The effects of the TOD coefficient on the  growth rate of  instability is demonstrated in \textbf{Fig.~10}, from which we discover that the value of $\Omega(z)$ increases with the value of $d_{3}(z)$. \textbf{Fig.~11} shows the characteristics of MI on the ($Q, n$) plane. It is found that the MI exists in the region $-2<Q<2$.
Hereby, we discover that the MI growth rate distribution is symmetric with respect to
\begin{equation}\label{MIE}
n=-\frac{d_{2}(z)}{6\,d_{3}(z)}\,,
\end{equation}
i.e.,
The red dashed line in Fig.~11 corresponds to a modulational stability (MS) region where
the growth rate is vanishing in the low perturbation frequency region.
More interestingly, using the rogue wave eigenvalue, one can find that the MS condition~(\ref{MIE}) is  consistent with the condition~(\ref{KZFC})
 which converts breathers into  nonlinear waves on constant backgrounds. Our finding suggests that the transition between breathers and nonlinear waves can occur in the MS region with the low frequency perturbations. Further, by comparison with \textbf{Fig.~11(a)}, we discover that the  lower value  of $d_{3}(z)$  corresponds to the lower value of  $n$. Therefore, the MS region moves down, which is displayed in \textbf{Fig.~11(b)}.

\vspace{5mm}
\noindent\textbf{\Large{\uppercase\expandafter{4}. Breather multiple births and Peregrine combs/walls}}

In this section, we will study the breather solution~(\ref{EXP-B1}) in detail, and describe its main properties when
the GVD coefficient  is of the form
\begin{equation}\label{AP}
d_{2}(z)=c\,R(z)=-1+d_{c}\,\cos(k_{c}\,z)\,,
\end{equation}
where $d_{c}$ denotes the amplitude of modulation and $k_{c}$ is spatial frequency.  Such periodic modulations have
realized  experimentally in Ref.~\cite{PM}.

The spatiotemporal characteristics of the breather multiple births described by the solution~(\ref{EXP-B1}) are illustrated in \textbf{Fig.~12}, i.e., the triplets structure with $d_{c}=2.5$ and the septuplets structure wtih $d_{c}=8.5$. These structures show multiple compression points, located at different values of $z$ and $t$. The number of the ABs in the breather multiple births  depends on the amplitude of the modulation but not on its wavelength, which controls their separation distance.  Increasing the value of $d_{c}$ will lead to the formation of ($3+4\,k$)-births for $k=2, 3,\ldots$. With the similar modulation parameters selected, these multiple births structures have been also reported in the vc-NLS equation, the vc-DNLS equation, and the vc-NLS-MB system. The difference is, however, that the phase shifts of the ABs on the sides of the center occur along $t$-direction because of the TOD effect (for example, see Fig.~12(a)), which don't exist in the vc-NLS equation without higher-order effect.
\begin{figure}[H]\centering
\subfigure[]{\includegraphics[width=100 bp]{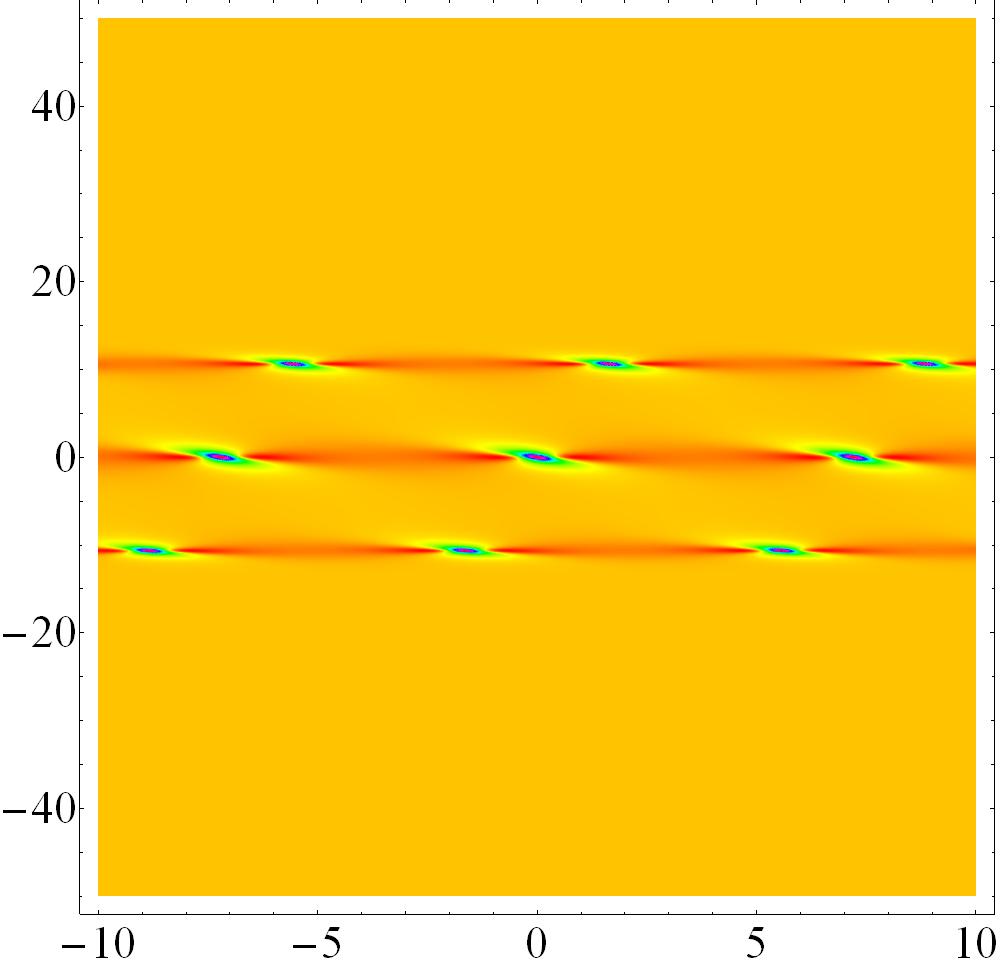}
\includegraphics[height=100 bp]{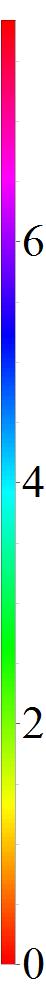}}
\qquad\qquad\qquad
\subfigure[]{\includegraphics[width=100 bp]{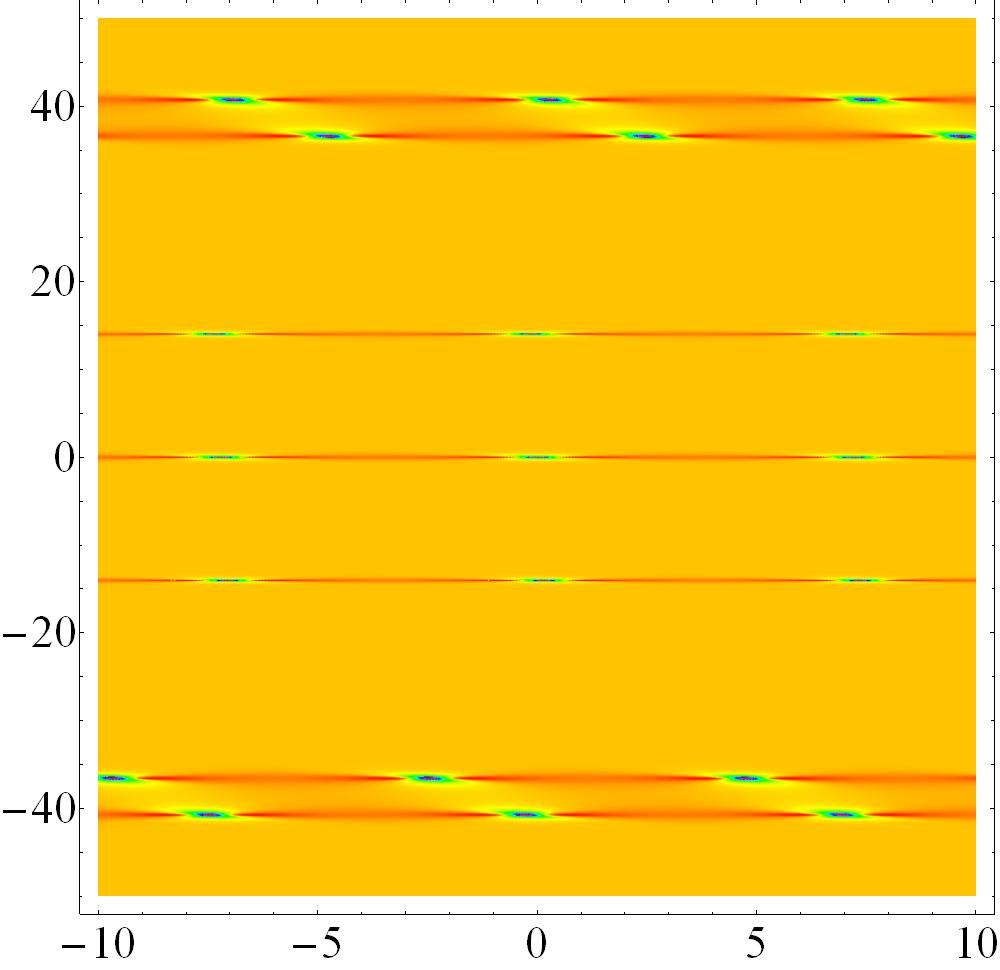}
\includegraphics[height=100 bp]{comb-xu0d9ji-nol1.jpg}}
\caption{\footnotesize The breather multiple births with $R(z)=d_{2}(z),\,k_{c}=0.2,\,d_{3}(z)=0.1,\,n=0,\,k_{1}=1,\,k_{2}=-1,\,\lambda_{1}=\lambda_{2}^{*}=0.9\,i\,$ and (a) 3-births with $d_{c}=2.5$, (b) 7-births with $d_{c}=8.5$. }
\end{figure}

\begin{figure}[H]\centering
\subfigure[]{\includegraphics[width=100 bp]{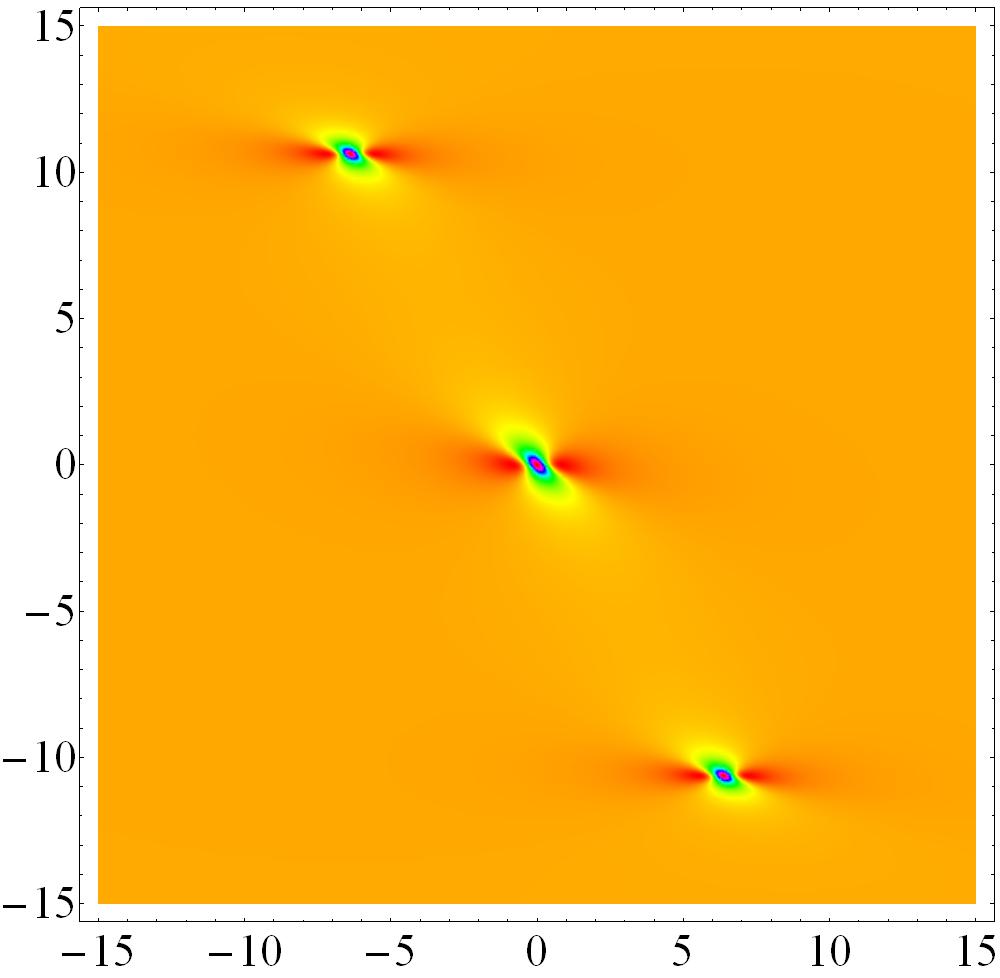}
\includegraphics[height=100 bp]{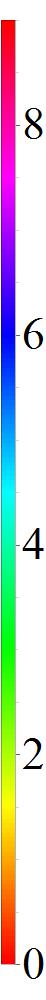}}\qquad\qquad
\subfigure[]{\includegraphics[width=150 bp]{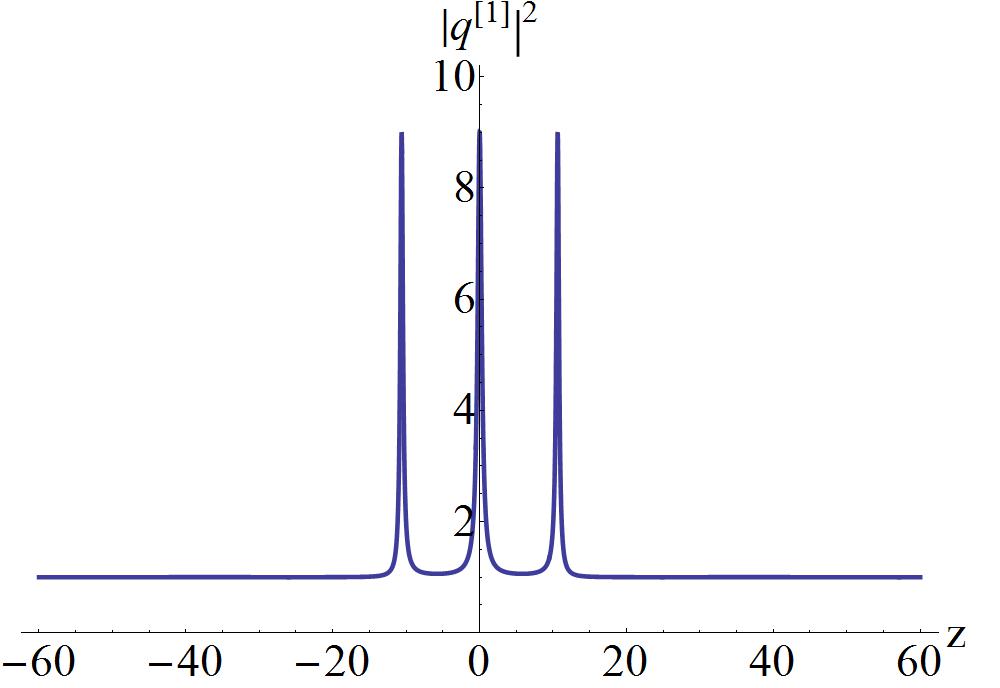}}\\
\subfigure[]{\includegraphics[width=100 bp]{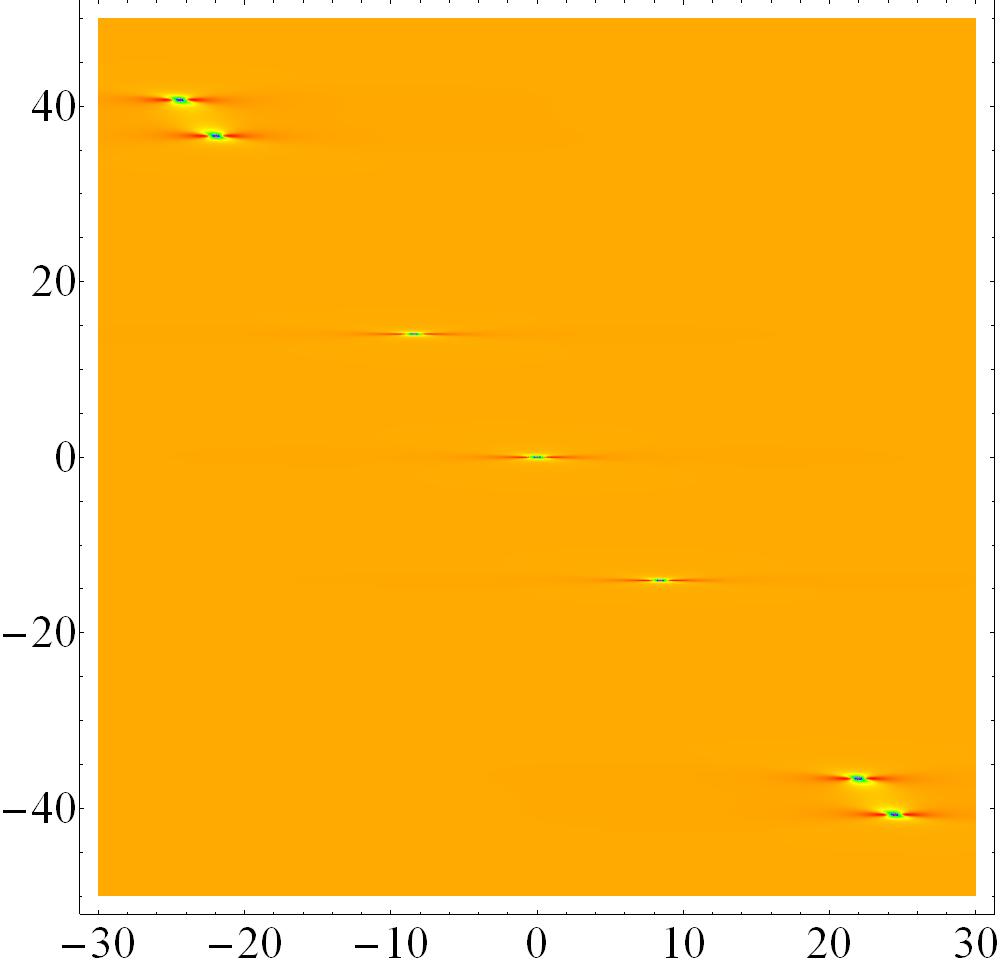}
\includegraphics[height=100 bp]{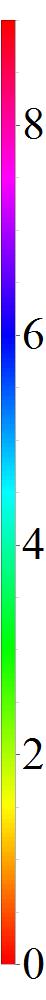}}
\caption{\footnotesize The Peregrine combs with $R(z)=d_{2}(z),\,k_{c}=0.2,\,d_{3}(z)=0.1,\,n=0,\,k_{1}=1,\,k_{2}=-1,\,\lambda_{1}=\lambda_{2}^{*}=-\frac{n}{2}+\,i\,$ and (a) triple tooth with $d_{c}=2.5$, (b) is the cross-sectional view of (a) along $t=-6\,d_{3}(z)z$. (c) seven tooth with $d_{c}=8.5$.}
\end{figure}

If $\lambda=\alpha+\beta\,i\rightarrow-\frac{n}{2}+\,i$, we can obtain another type of multiple compression point structure, namely the Peregrine combs that are the limiting case of the breather multiple births.  The Peregrine comb solution can be given by
\begin{equation}\label{COMP}
q_{comb}(z,t)=e^{i\rho}\Big(c(z)-2\frac{(i+2\,Z_{c})^{2}+4\,M^{2}}{1+4\,Z_{c}^{2}+4\,M^{2}}\Big)\,,
\end{equation}
with
 \begin{equation}
\begin{aligned}\label{AB113245}
&Z_{c}=-z\,+\frac{d_{c}}{k_{c}}\sin(k_{c}\,z)\,,\nonumber
&M=t+6\,z\,d_{3}(z)\,.\nonumber
\end{aligned}
\end{equation}
From the above expression, we see that the Peregrine comb solution~(\ref{COMP}) includes the gain or loss $\Gamma(z)$ that controls the amplitude,  and   TOD coefficient $d_{3}(z)$ that affect the spatial-temporal distribution.
The Peregrine combs were first found in the vc-NLS equation~\cite{PCC}, and then were also found in the vc-coupled Hirota equations~\cite{PCC1}. These phenomena do not take place in the standard AB or PS without variable dispersion, which contain only one compression point.
The maximum value of this wave's amplitude is obtained at $t=-6\,d_{3}(z)z$ and is given as follow
\begin{equation}\label{AB3fg455}
|q_{comb}|_{max}^{2}(z)=|q_{comb}(z,-6\,d_{3}(z)z)|^{2}=\frac{(1-4\,Z_{c}^{2})4\,c(z)}{1+4\,Z_{c}^{2}}+c(z)^{2}+4\,,
\end{equation}
with
 \begin{equation}
\begin{aligned}\label{AB113245}
&Z_{c}=-z\,(1-\frac{d_{c}}{k_{c}\,z}\sin(k_{c}\,z))\,.
\end{aligned}
\end{equation}
The GVD  coefficient $d_{2}(z)$ or TOD coefficient $d_{3}(z)$ has on effect on the  maximum
intensity of the Peregrine comb, which is related to the gain or loss coefficient $\Gamma(z)$.
From the equation~(\ref{AB3fg455}), we find that $|q_{comb}|_{max}^{2}(z)$ reaches its maximal value $(c(z)+2)^{2}$ at $Z_{c}=0$.
This means that the compression points of the Peregrine comb are located at $z_{0}=0$, $z_{i}$ ($i=1, 2,\ldots$), where  $z_{i}$ satisfy the following equaiton
 \begin{equation}
\begin{aligned}\label{AB11rt}
&\frac{\sin(k_{c}z_{i})}{k_{c}z_{i}}=\frac{1}{d_{c}}\,.
\end{aligned}
\end{equation}
Eq.~(\ref{AB11rt}) shows that the larger the value of $d_{c}$ is, the more compression points will be.
For example, \textbf{Fig.~13(a)}  displays a Peregrine comb  with triple tooth of Eq.~(\ref{vcHirota}) with $d_{c}=2.5$, and \textbf{Fig.~13(c)} is plotted for a Peregrine comb with seven tooth with $d_{c}=8.5$.  The detailed generation process of Peregrine combs with seven tooth can refer to the explanations for the vc-NLS equation in Ref.~\cite{PCC}.

\begin{figure}[H]\centering
\subfigure[]{\includegraphics[width=100 bp]{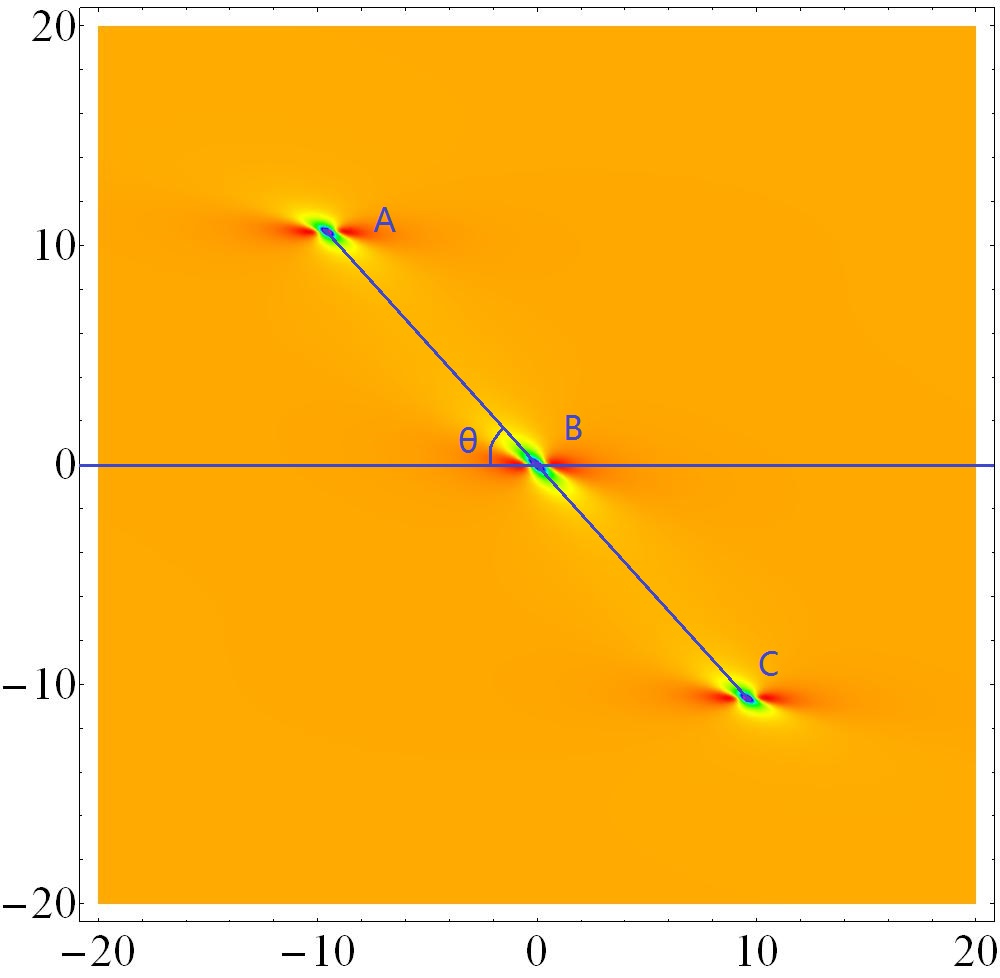}
\includegraphics[height=100 bp]{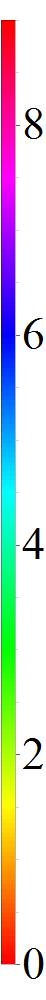}}
\quad\quad\quad
\subfigure[]{\includegraphics[width=100 bp]{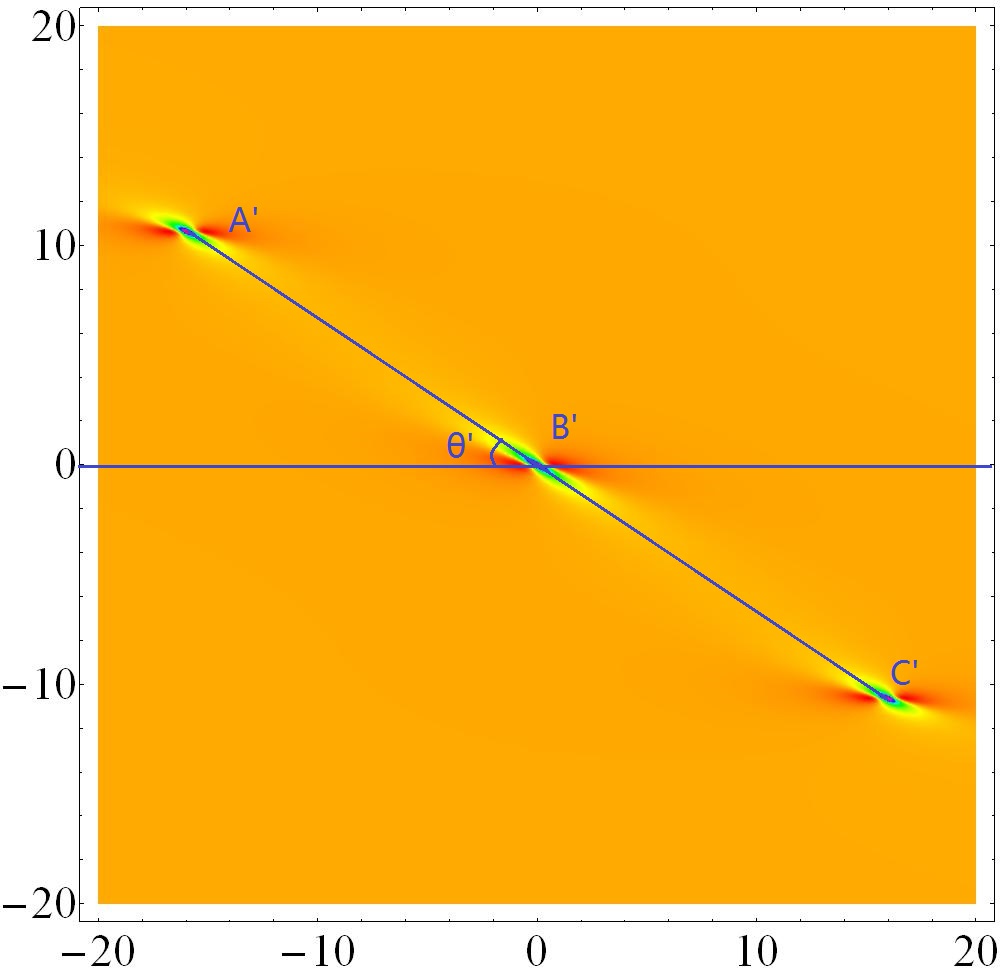}
\includegraphics[height=100 bp]{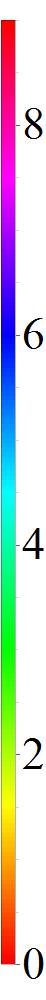}}

\caption{\footnotesize The  effect of $d_{3}$ on the Peregrine comb with $R(z)=d_{2}(z),\,d_{c}=2.5,\,k_{c}=0.2,\,n=0,\,k_{1}=k_{2}=1,\,\lambda_{1}=\lambda_{2}^{*}=-\frac{n}{2}+\,i\,$ and (a) $d_{3}(z)=0.15$, (b) $d_{3}(z)=0.25$. }
\end{figure}

In order to explore the dynamics of the Peregrine comb, we calculate some physical quantities of the Peregrine comb  analytically. The maximum points,  intersection angle $\theta$ and the distance between \emph{A} and \emph{B} can be presented as follows
 \begin{equation}
\begin{aligned}\label{ABss11}
(0,0), \qquad (-\frac{\arccos(\frac{1}{d_{c}})}{k_{c}},\frac{6\,d_{3}(z)\arccos(\frac{1}{d_{c}})}{k_{c}}),\qquad(\frac{\arccos(\frac{1}{d_{c}})}{k_{c}},-\frac{6\,d_{3}(z)\arccos(\frac{1}{d_{c}})}{k_{c}})\,,
\end{aligned}
\end{equation}
 \begin{equation}
\begin{aligned}\label{AB111111}
\theta=\arctan(-\frac{1}{6\,d_{3}(z)})\,,
\end{aligned}
\end{equation}
 \begin{equation}
\begin{aligned}\label{AB11143111}
D_{AB}=\sqrt{1+36\,d_{3}^{2}(z)}\frac{\arccos(\frac{1}{d_{c}})}{k_{c}} \,.
\end{aligned}
\end{equation}
Bases on the above analytic expressions, we find that the TOD effect $d_{3}(z)$  plays an important role in the spatiotemporal characteristics of the Peregrine comb. Increasing the values of $d_{3}(z)$ will increase the distance between $A$ and $B$ while decrease the intersection angle $\theta$, which is shown in\textbf{ Fig.~14}. If $d_{3}(z)$ is equal to zero, the Peregrine comb in Fig.~14 will degenerate into the case in Ref.~\cite{PCC}.

In order to reveal further the  characteristics of the Peregrine comb, we introduce the energy of light pulse against the plane-wave background with the
form~\cite{PCC,ENG}
\begin{equation}\begin{aligned}\label{AB3455}
&\Delta\,I_{c}(z, t)=|q_{PS}(z,t)|^{2}-c(z)^{2}=4\frac{\,F_{+}\,F_{-}-\,c(z)\,K_{+}\,K_{-}}{(1+4\,Z^{2}+4(t+6\,z\,d_{3}(z))^{2})^{2}}\,,
\end{aligned}\end{equation}
with
 \begin{equation}
\begin{aligned}\label{AB113sd5}
&F_{-}=4\,t^{2}+4\,Z^{2}+(1-12\,z\,d_{3}(z))^{2}+t(-4+48\,z\,d_{3}(z))\,,\\
&F_{+}=4\,t^{2}+4\,Z^{2}+(1+12\,z\,d_{3}(z))^{2}+t(4+48\,z\,d_{3}(z))\,,\\
&K_{-}=-1+4\,Z^{2}+4\,(t+6\,z\,d_{3}(z))^{2}\,,\\
&K_{+}=1+4\,Z^{2}+4\,(t+6\,z\,d_{3}(z))^{2}\,.\\
\end{aligned}
\end{equation}
One can easily check
\begin{equation}\begin{aligned}\label{AB3455}
&\int_{-\infty}^{\infty}\Delta\,I_{c}(z,t)dt=0\,,
\end{aligned}\end{equation}
for all values of $z$. This implies that the energy of the pump is preserved along the fiber,
in spite of periodic modulation characteristics and TOD effect added. In addition, Eq.~(\ref{AB113sd5}) also reflects the fact that
 the light intensity of the Peregrine comb will be
sometimes higher and sometimes lower than the background
intensity, as shown in \textbf{Fig.~15}. Moreover, we can calculate the energy of the Peregrine pulse
\begin{equation}\begin{aligned}\label{AB3455}
E_{pluse}(z)=&\int_{-\infty}^{\infty}|q_{comb}(t,z)-q_{comb}(\pm\infty,z)|^{2}dt=\frac{4\,\pi}{\sqrt{1+4Z_{c}^{2}}}\,,
\end{aligned}\end{equation}
which indicates that the energy of the pulse is maximal at the compression
points $Z_{c} = 0$ (also see \textbf{Fig.~16}).
\begin{figure}[H]\centering
\includegraphics[width=200 bp]{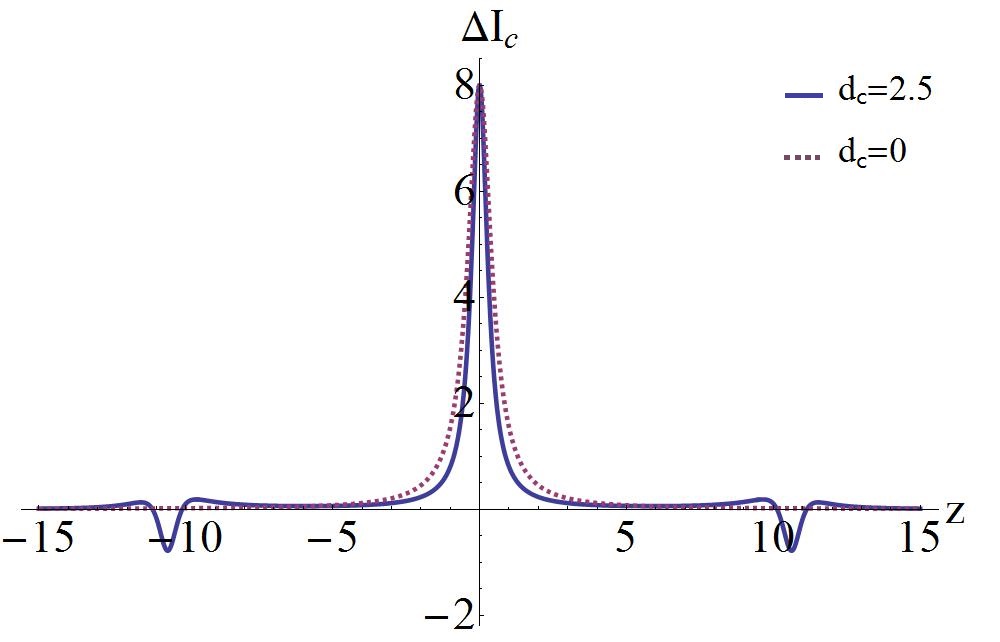}
\caption{\footnotesize The distribution of the difference between the light intensities of the PS and the CW background at $t=0$ given by Eq.~(35) with $k_{c}=0.2,\,d_{3}=0.02,\,d_{c}=0$ (dashed line), and $d_{c}=2.5$ (solid line).}
\end{figure}
\begin{figure}[H]\centering
\includegraphics[width=200 bp]{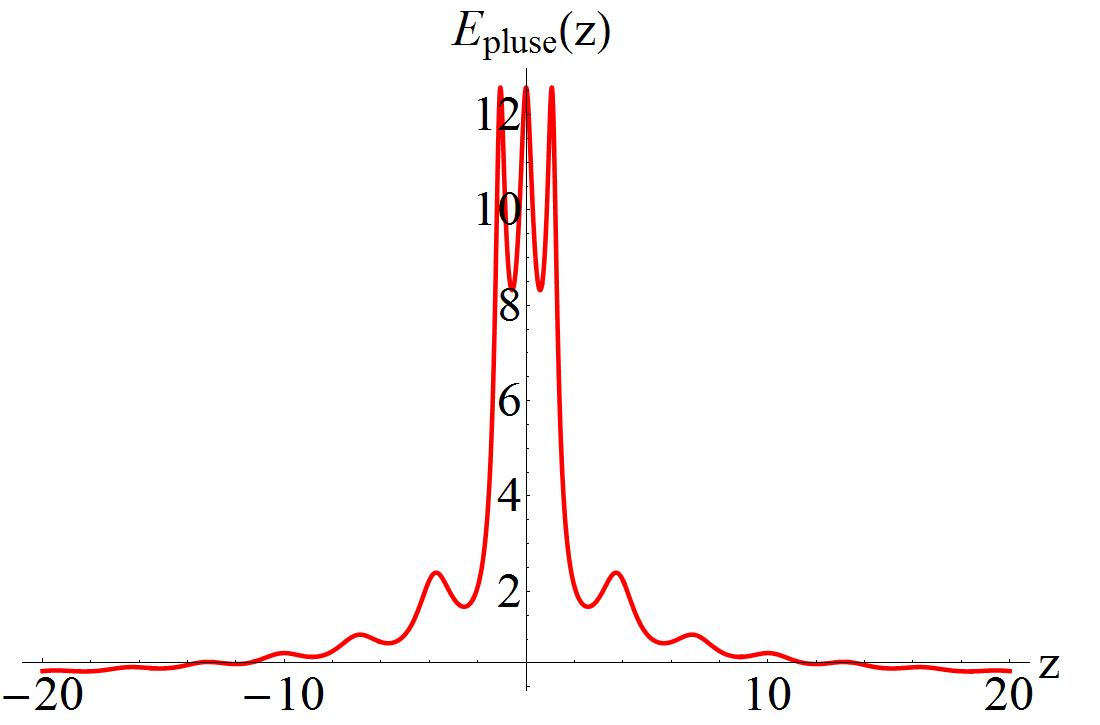}
\caption{\footnotesize Pulse energy $E_{pulse}(z)$ with $k_{c}=2,\,d_{c}=2.5.$}
\end{figure}

Next, we consider another special case of  the generalized PS solution, namely the Peregrine walls.
From the equation~(\ref{ABss11}), we note that the amplitude of modulation $d_{c}$ and spatial frequency $k_{c}$ also affect the positions of two of maximum points, i.e., $(\mp\frac{\arccos(\frac{1}{d_{c}})}{k_{c}},\pm  \frac{6\, d_{3}(z)\arccos(\frac{1}{d_{c}})}{k_{c}})$. In particular, we consider a  special case in which the amplitude of modulation $d_{c}$ is set to be 1. This will result in the case $\arccos(\frac{1}{d_{c}})=0$. Thus, these two maximum points will be shifted to the origin and these three PSs will be aggregated together. In this case, the Peregrine comb solution~(\ref{AB3455}) turns into the form of
\begin{equation}\label{WALL}
|q_{wall}(z,t)|^{2}=(c(z)-2)^{2}+\frac{M\,(8+9\,M)}{W^{2}}+\frac{2\,(-6\,M+(4+3\,M)c(z))}{W}\,,
\end{equation}
with
 \begin{equation}
\begin{aligned}\label{AB113245}
&W=(1+M+4\,z^{2})+4\frac{\sin(k_c z)}{k_{c}^{2}}(\sin(k_c z)-2\,k_c z)\,.\nonumber
\end{aligned}
\end{equation}
The maximum value of this wave's amplitude is obtained at $t=-6\,d_{3}(z)z$ and is given by
\begin{equation}\label{AB3fg455}
|q_{wall}|_{max}^{2}(z)=|q_{wall}(z,-6\,d_{3}(z)z)|^{2}=(c(z)-2)^{2}+\frac{8\,c(z)}{W_0}\,,
\end{equation}
with
 \begin{equation}
\begin{aligned}\label{AB113245}
&W_0=(1+4\,z^{2})+4\frac{\sin(k_c z)}{k_{c}^{2}}(\sin(k_c z)-2\,k_c z)\,.\nonumber
\end{aligned}
\end{equation}
As depicted in \textbf{Fig.~17}, the Peregrine comb is converted into a {Peregrine wall} when the three maximum points have the same  coordinate ($0, 0$). Generally, the fusion of three PSs will produce a second-order  rogue wave with higher amplitude in the nonlinear equation of evolutions with constant coefficients. However, the variable coefficients provide much richer patterns. The Peregrine wall can be seen as an intermediate state between the rogue wave and W-shaped soliton because rogue wave has a shorter life while W-shaped soliton has a long one. Such structure looks like a quasi-trapezoidal in shape. The values of the hump and valleys of the Peregrine wall, respectively, are equal to $(c(z)+2)^{2}$ and 0. To illustrate the effect of TOD coefficient on the Peregrine wall, we plot \textbf{Fig.~18}. It is observed that the length of the wave increases with growing value of $d_{3}(z)$, and the depth and angle decreases. In fact, similar to the Peregrine comb, the Peregrine wall can also be viewed as the limiting case of a breath-type wall that is described in \textbf{Fig.~19}. On the other hand, to illustrate how to build a  Peregrine wall from a PS, we plot \textbf{Fig.~20} with different values of $d_{c}$.
As $d_{c}$ increases, the lifetime of the PS gets much longer. When $d_{c}$ is equal to 1, the PS eventually becomes a Peregrine wall. When $d_{c}$ is greater than 1, the PS will show
multiple compression points, i.e., it  turns into the Peregrine comb~\cite{PCC}.

\begin{figure}[H]\centering
\subfigure[]{\includegraphics[width=150 bp]{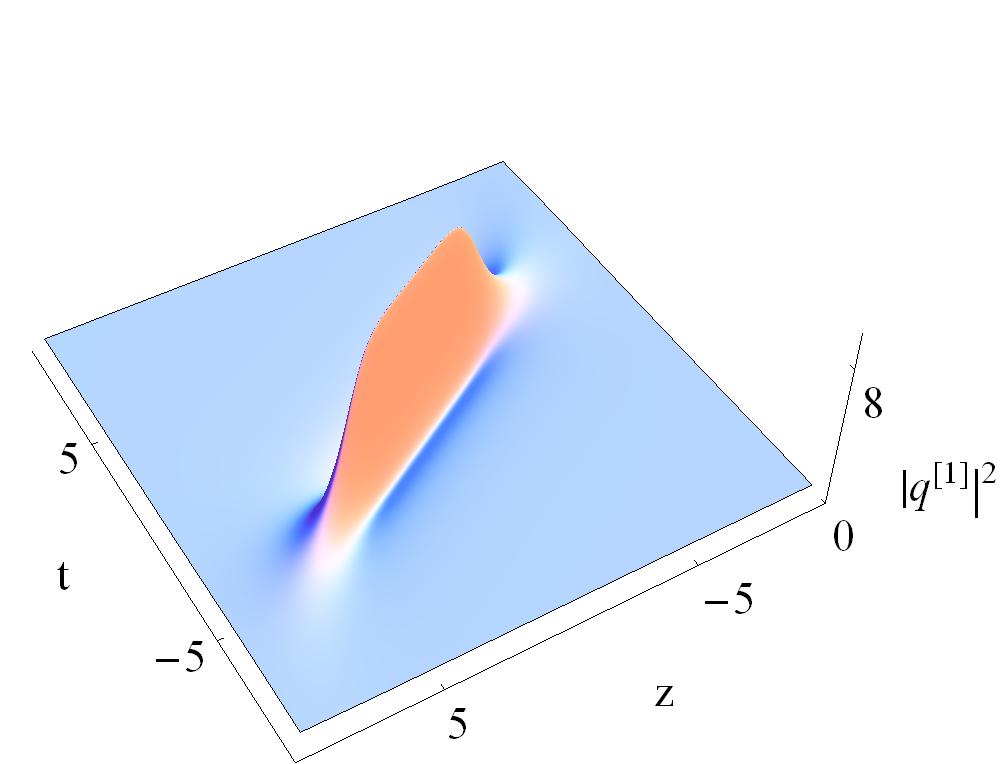}}\qquad\qquad
\subfigure[]{\includegraphics[width=100 bp]{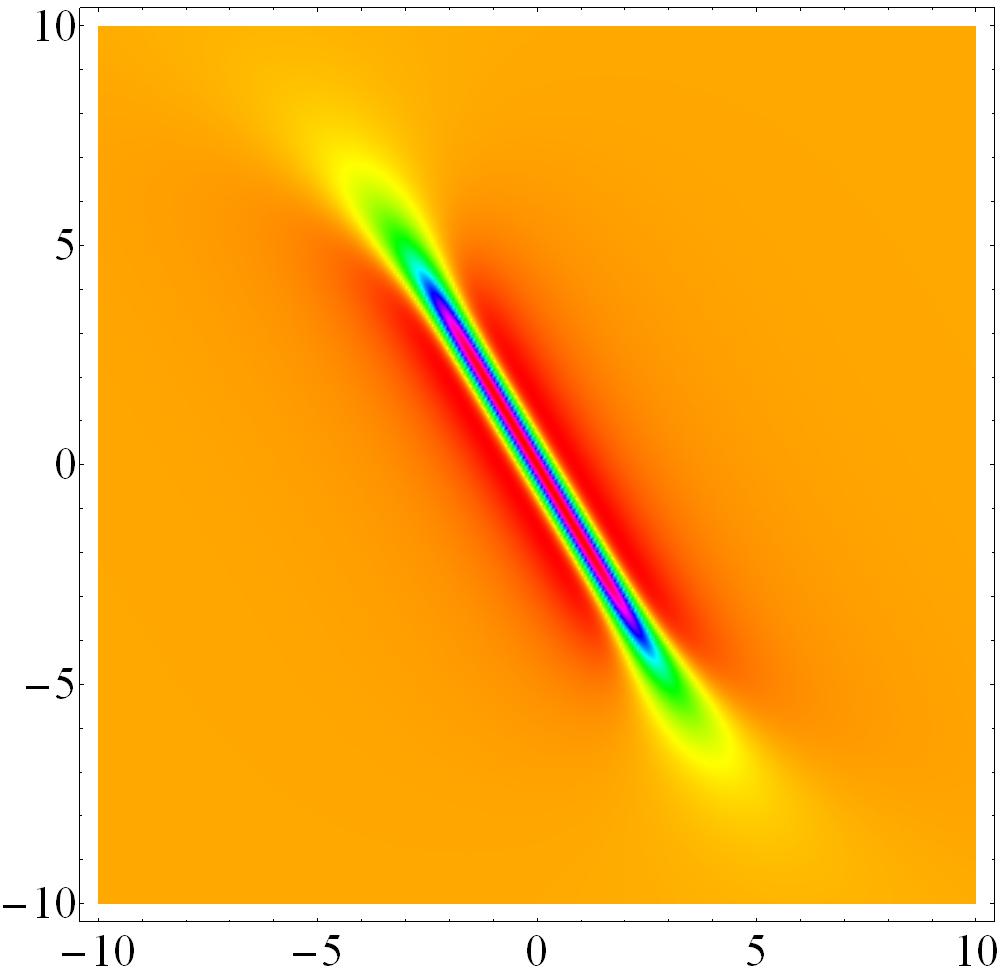}
\includegraphics[height=100 bp]{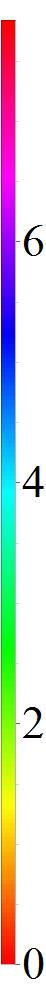}}
\caption{\footnotesize The Peregrine wall with $R(z)=d_{2}(z),\,k_{c}=0.2,\,d_{3}(z)=0.1,\,n=0,\,k_{1}=1,\,k_{2}=-1,\,\lambda_{1}=\lambda_{2}^{*}=-\frac{n}{2}+\,i\,$ and $d_{c}=1$. (b) is the density plot of (a). }
\end{figure}
\begin{figure}[H]\centering
\subfigure[]{\includegraphics[width=150 bp]{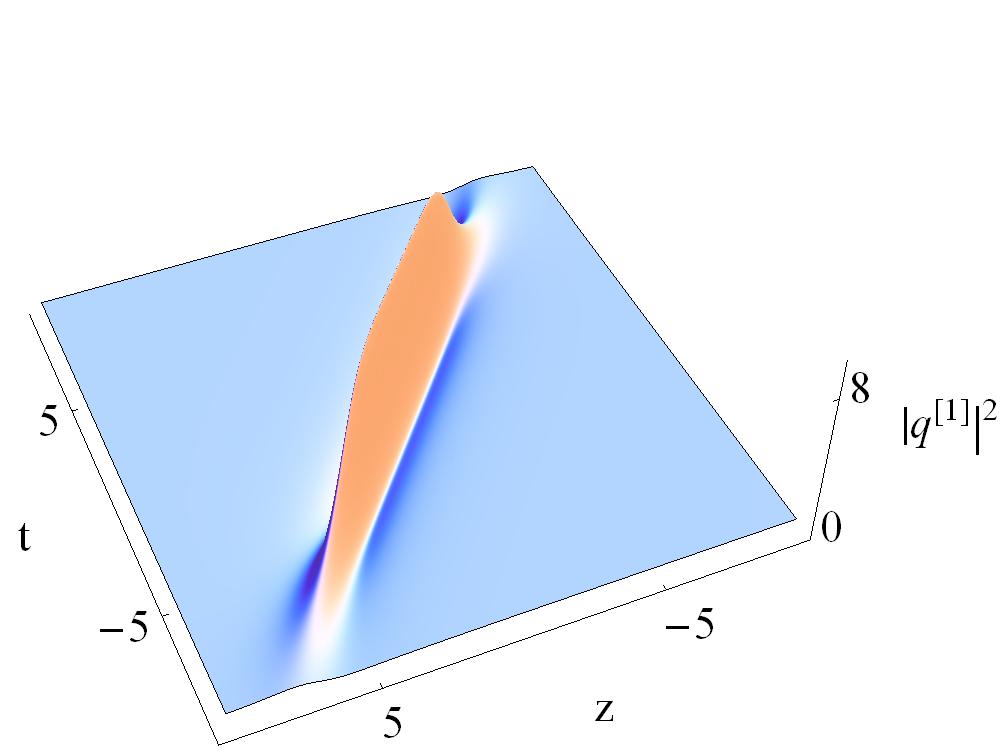}}\qquad\qquad
\subfigure[]{\includegraphics[width=100 bp]{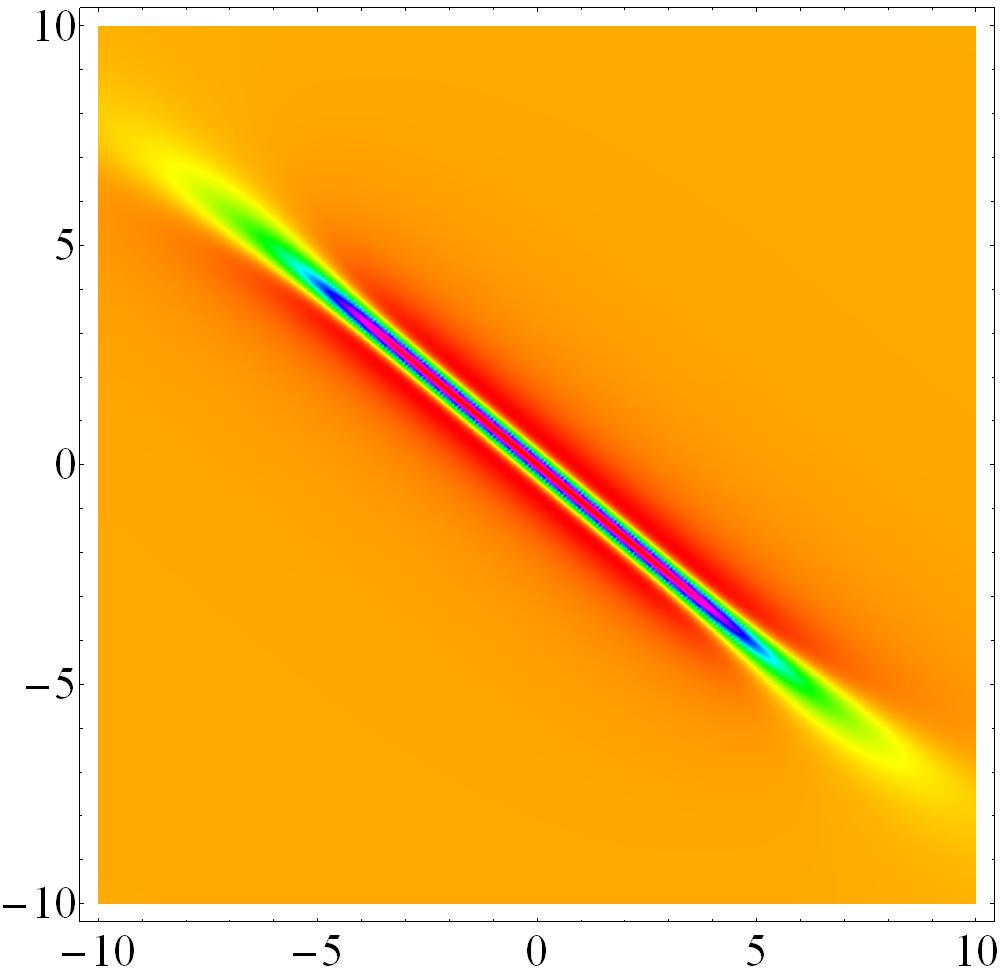}
\includegraphics[height=100 bp]{shuzi-huxizi-qiang-D1.jpg}}
\caption{\footnotesize The effect of $d_{3}(z)$ on the Peregrine wall with  $R(z)=d_{2}(z),\,k_{c}=0.2,\,d_{3}(z)=0.2,\,n=0,\,k_{1}=1,\,k_{2}=-1,\,\lambda_{1}=\lambda_{2}^{*}=-\frac{n}{2}+\,i\,$ and $d_{c}=1$. (b) is the density plot of (a). }
\end{figure}

\begin{figure}[H]\centering
\subfigure[]{\includegraphics[width=150 bp]{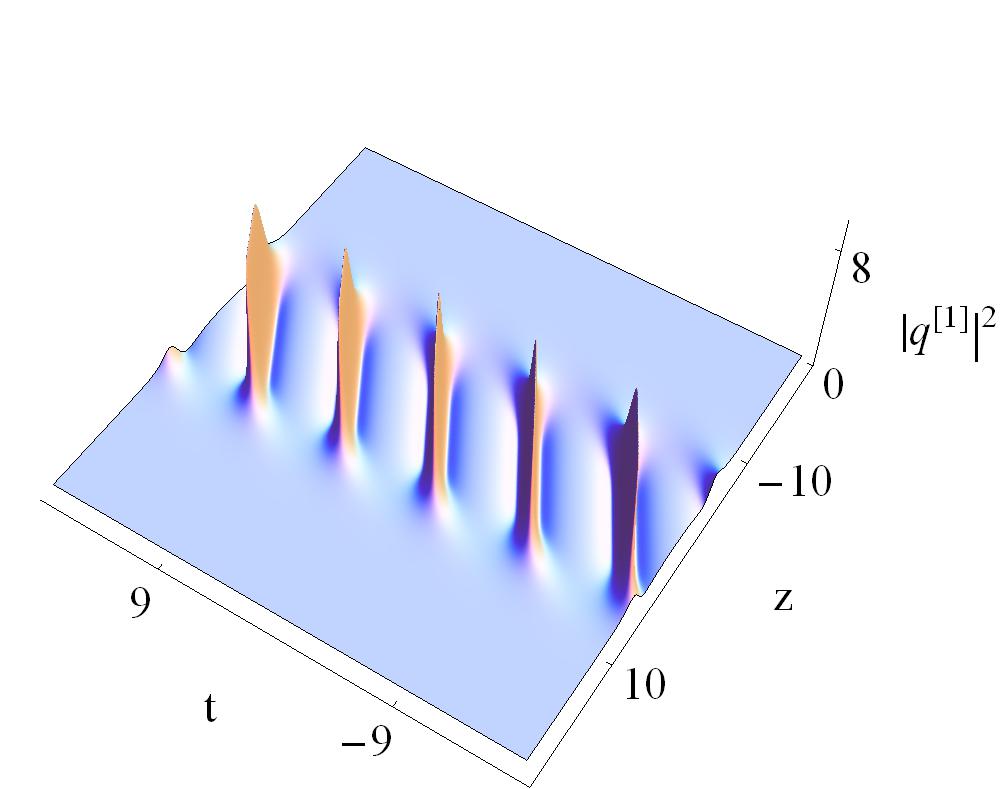}}\qquad\qquad
\subfigure[]{\includegraphics[width=100 bp]{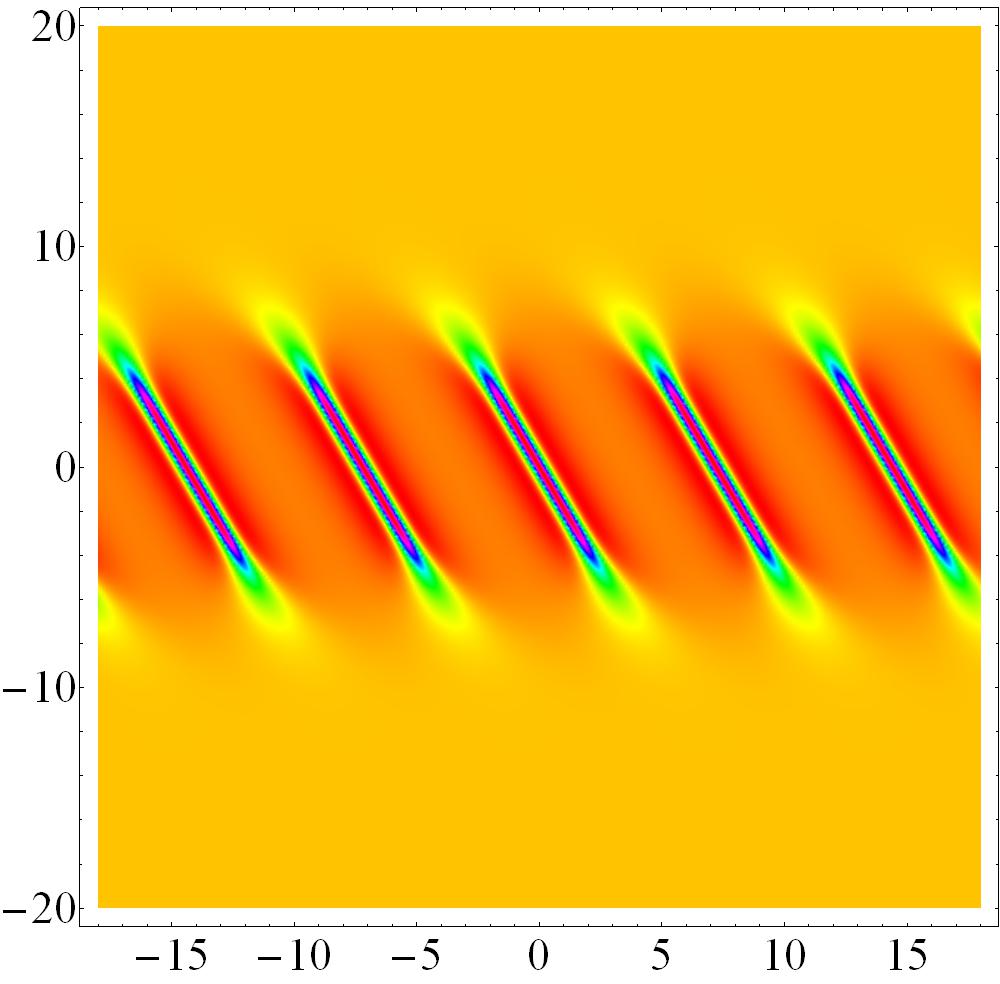}
\includegraphics[height=100 bp]{shuzi-huxizi-qiang-D1.jpg}}
\caption{\footnotesize The  breath-type wall with $R(z)=d_{2}(z),\,k_{c}=0.2,\,d_{3}(z)=0.1,\,n=0,\,k_{1}=1,\,k_{2}=-1,\,\lambda_{1}=\lambda_{2}^{*}=-\frac{n}{2}+0.9\,i\,$ and $d_{c}=1$. (b) is a density plot of (a).}
\end{figure}

\begin{figure}[H]\centering
\includegraphics[width=200 bp]{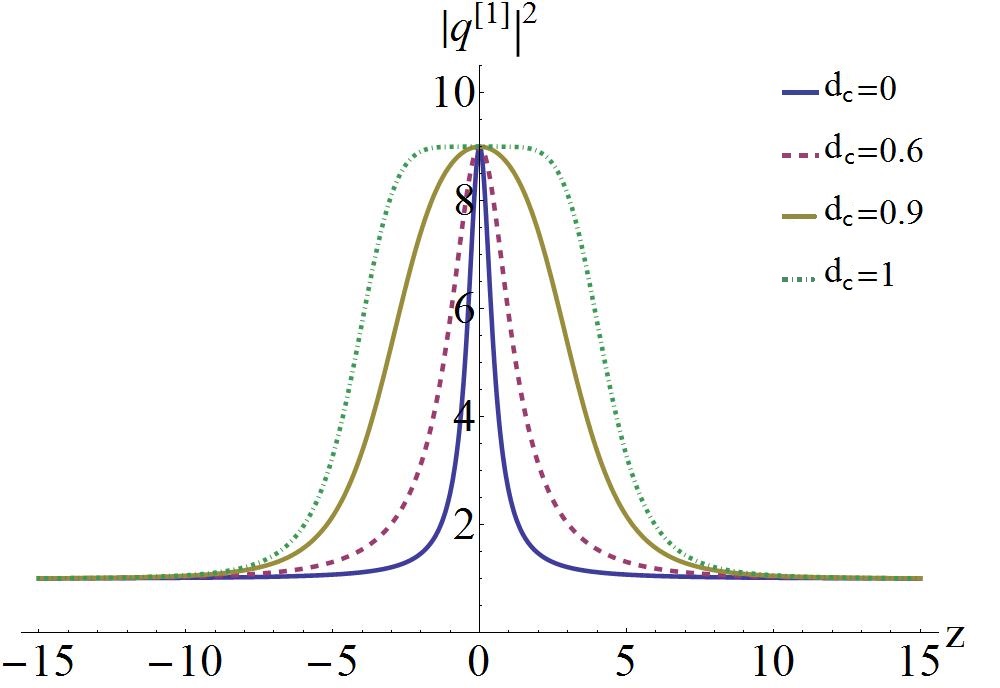}
\caption{\footnotesize Graph of $|q_{wall}(z,-6\,d_{3}(z)z)|^{2}$ for value of $d_{c}$ with $k_{c}=0.2,\,d_{3}=0.1$ shows the evolution process from rogue wave to the comb.}
\end{figure}
The frequency spectrum of the Peregrine comb in Eq.~(\ref{COMP}) can be given by
\begin{equation}\begin{aligned}\label{Fr}
F(\omega,z)&=\frac{1}{\sqrt{2\pi}}\int_{-\infty}^{\infty}q_{comb}(z,t)e^{i\,\omega\,t}dt\\
&=\sqrt{2\pi}\Big(\frac{1-2\,i\,Z_{c}}{\sqrt{1+4\,Z_{c}^{2}}}e^{-\frac{i\,|\omega|}{2}(12\,d_{3}(z)\,z+\sqrt{-1-4Z_{c}^{2}})}-\delta(\omega)\Big)\,,
\end{aligned}\end{equation}
where the Dirac delta function $\delta(\omega)$ originates from the finite
background level. The modulus of this spectrum is given by
\begin{equation}\begin{aligned}\label{F1r}
|F(\omega,z)|&=\sqrt{2\pi}e^{-\frac{|\omega|}{2}\sqrt{1+4\,Z_{c}^{2}}}\,.
\end{aligned}\end{equation}

 It is well known that the Peregrine rogue wave spectrum
features a triangular shape and gets dramatically broadened
at the maximally compressed peak. To end this section, we consider the
spectral property of the Peregrine combs and walls. From \textbf{Fig.~21(a)},  we observe that the rogue wave in Eq.~(\ref{vcHirota}) begin with
narrow spectral components as the constant-coefficient ones, but spreads and shrinks during
the evolution along the fiber, and eventually restore its initial shape. Each nonlinear spreading in the spectrum is
related to a corresponding maximal compression point.
Therefore, the number of spectral components increases with the amplitude of modulation, which is displayed in \textbf{Fig.~21(b)}. \textbf{Fig.~21(c)} shows the spectrum of the Peregrine wall. It is observed that the width of nonlinear spreading are obviously greater than that of the Peregrine comb.
\begin{figure}[H]\centering
\subfigure[]{\includegraphics[width=130 bp]{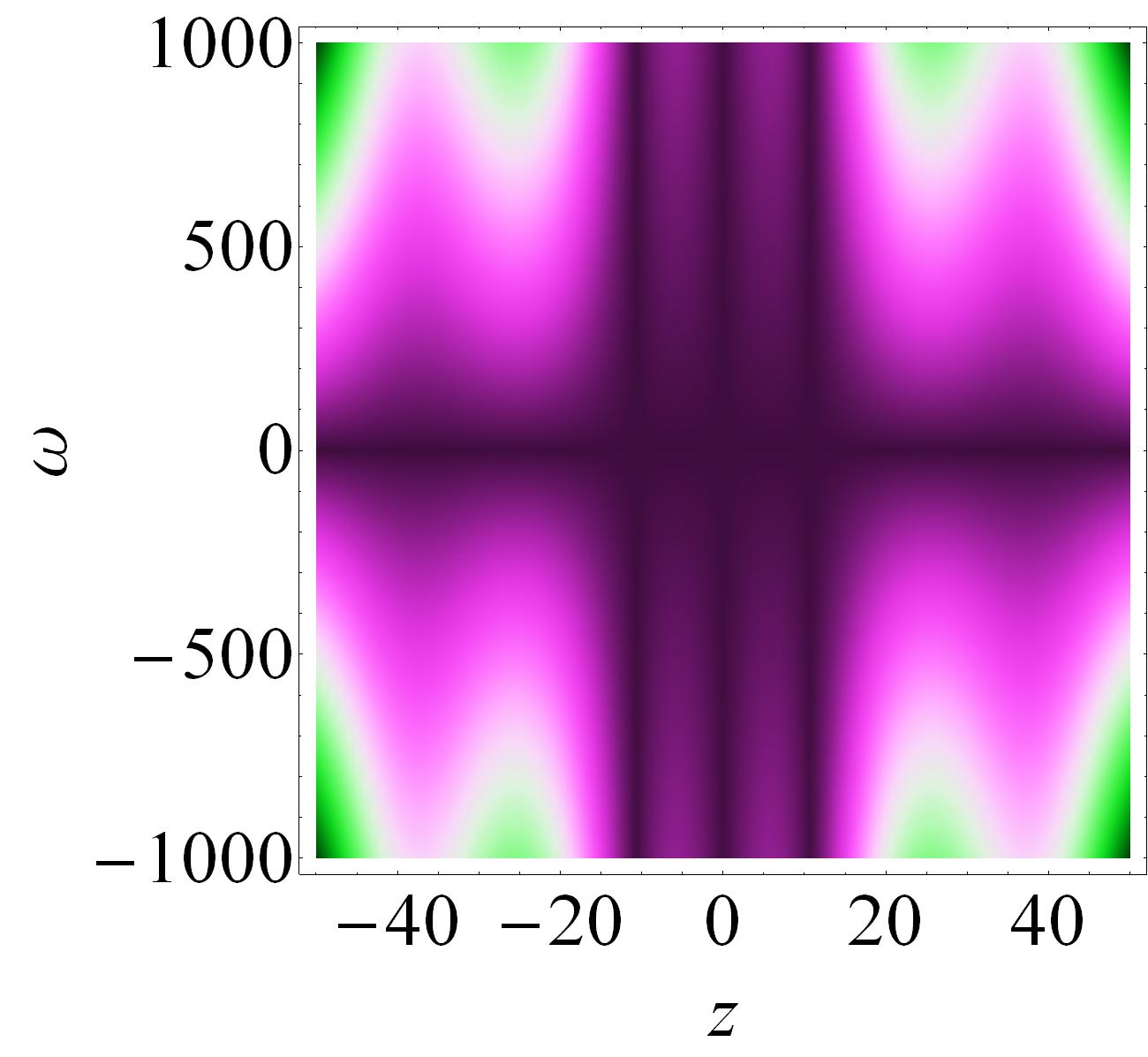}
\includegraphics[height=120 bp]{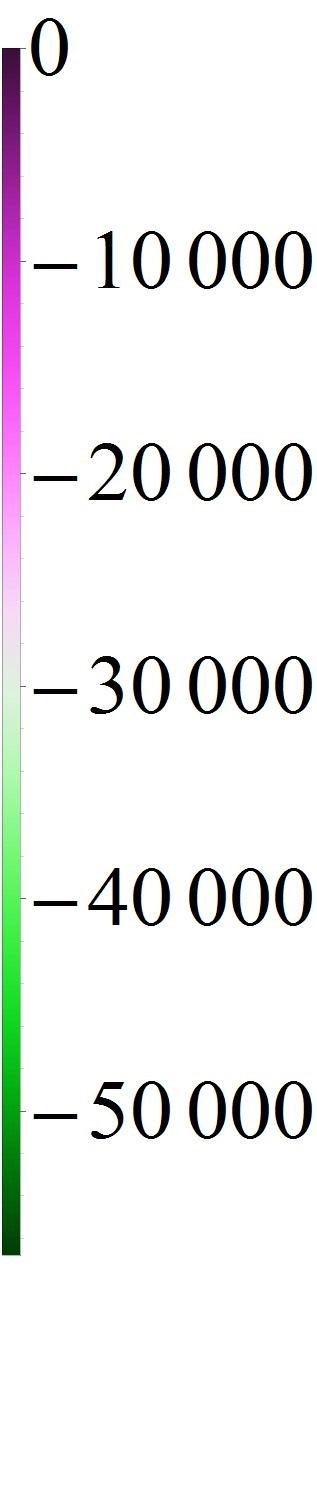}}
\subfigure[]{\includegraphics[width=130 bp]{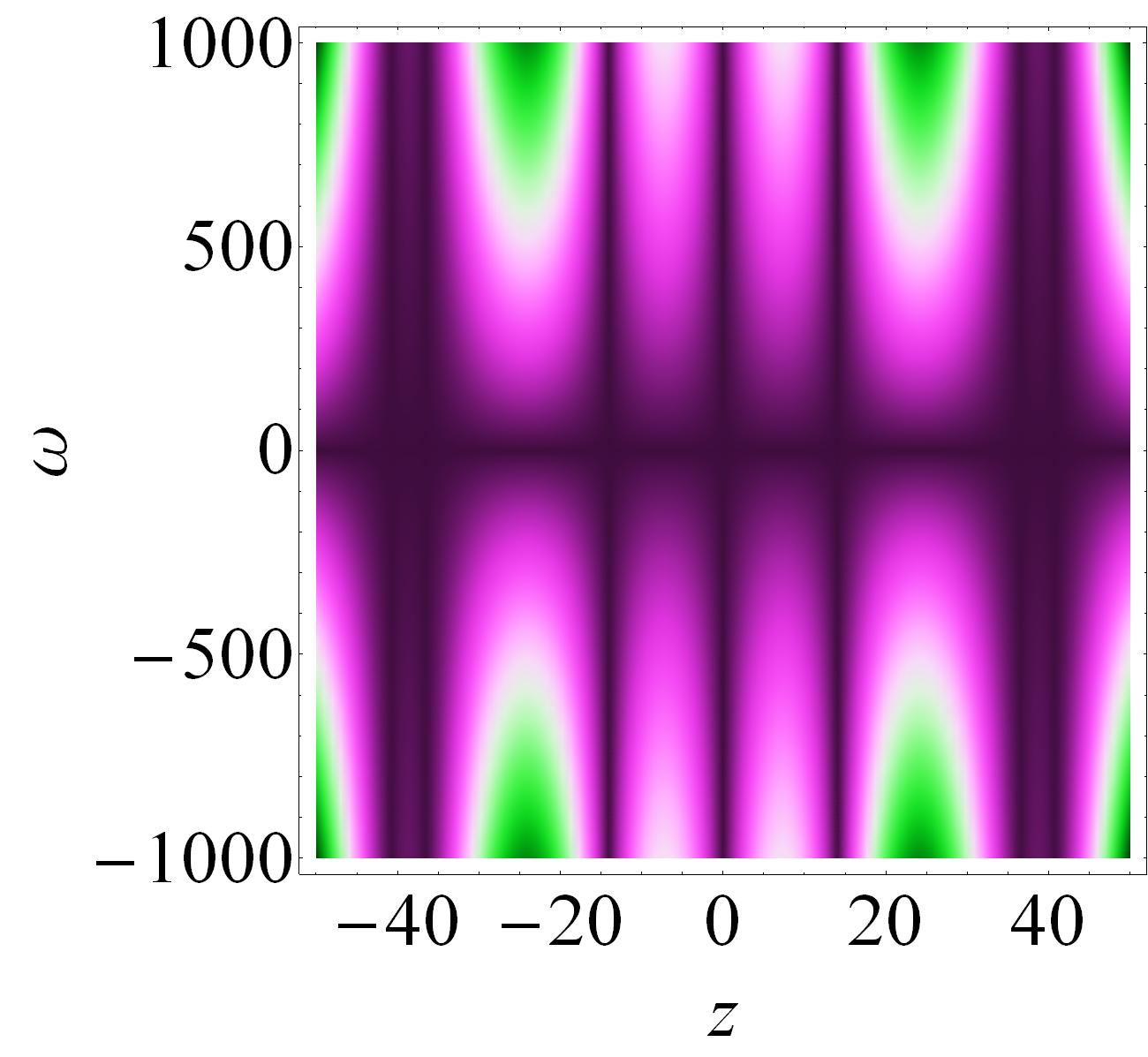}
\includegraphics[height=120 bp]{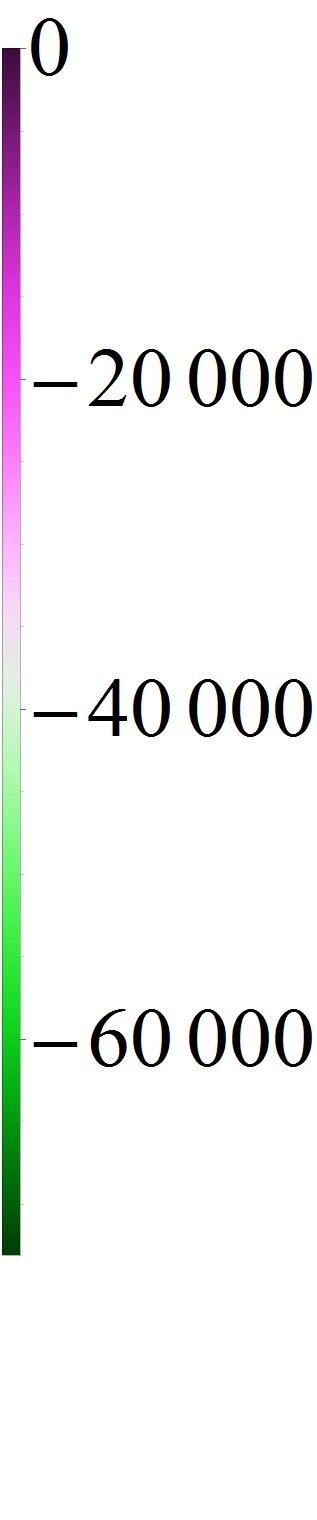}}
\subfigure[]{\includegraphics[width=130 bp]{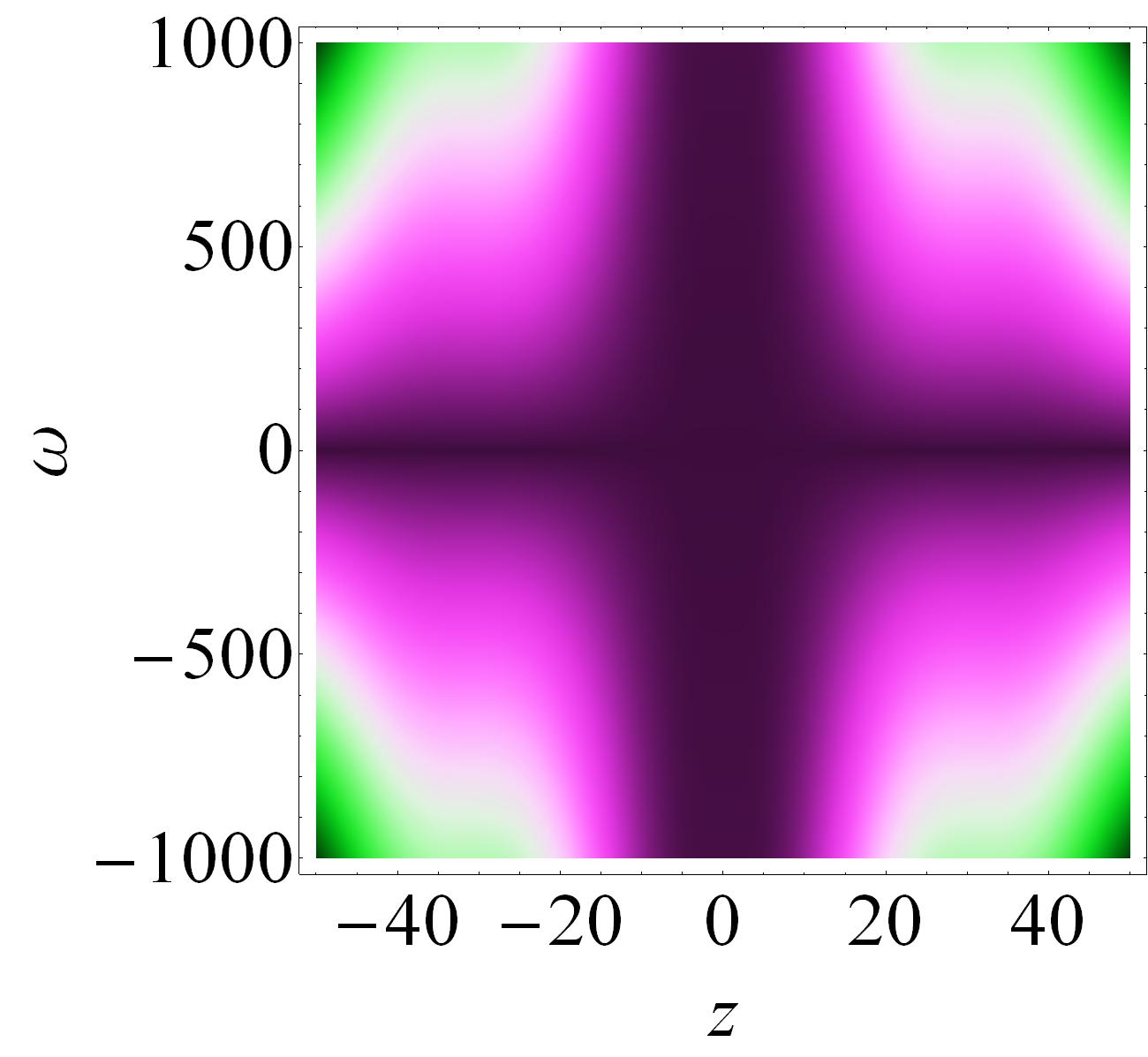}
\includegraphics[height=120 bp]{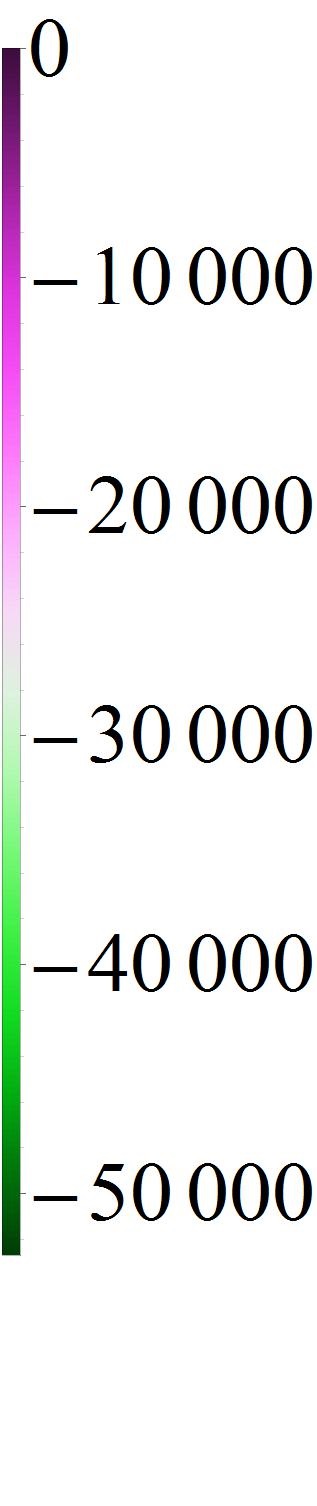}}
\caption{\footnotesize The spectrum of the Peregrine combs and Peregrine wall log scale, namely, $\log|F(w, z)|$ $k_{c}=0.2,\,d_{3}(z)=0.1$ and (a) For the triple tooth comb with $d_{c}=2.5$, (b) For the seven tooth comb with $d_{c}=8.5$, (c) For the Peregrine wall with $d_{c}=1$. }
\end{figure}

\vspace{5mm}
\noindent\textbf{\Large{\uppercase\expandafter{5}. Conclusions}}

We have carried out the analytical investigations on the vc-NLS equation with higher-order effects. The breathers solution have been shown that they can be converted into the multi-peak solitons (single and double main peaks), antidark soliton, periodic wave and W-shaped soliton. The transition condition depending on the eigenvalue, GVD coefficient and
TOD coefficient has been given analytically. We have demonstrated that different types of nonlinear waves can coexist and interact with each other elastically. We have further revealed the effects of the variable coefficients on the multi-peak solitons: (1) The GVD coefficient controls the number of peaks of the wave; (2) The TOD coefficient  accounts for  the compressed effects of the wave; (3) The gain or loss coefficient is responsible for the amplitude of the wave. The transition  between breathers and nonlinear waves occurs in the MS region with the low frequency perturbations. We have shown that, under the suitable periodic modulations, the Peregrine combs and Peregrine walls are formed. We have discovered that the TOD coefficient has influence on the spatiotemporal characteristics of the  Peregrine combs and Peregrine walls. Our results could provide certain theoretical assistance to the experimental control
and manipulation of generalized rogue wave dynamics in inhomogeneous fiber.

\vspace{5mm}
\noindent\textbf{\Large{ Acknowledgements}}

We express our sincere thanks to all the members of our discussion group for their valuable comments. This work has been supported by the National Natural Science Foundation of China under Grant (Nos.~11305060 and 61505054), by the Fundamental Research Funds of the Central Universities (No.~2015ZD16), and by the Innovative Talents Scheme of North China Electric Power University.

\vspace{5mm}
\noindent\textbf{\Large{ Appendix}}

 With the Ablowitz-Kaup-Newell-Segur formalism, the Lax pair associated with Eq.~(\ref{vcHirota}) can be
written as~\cite{LL}
\renewcommand\theequation{A1}
\begin{equation}
\Phi_t=U\,\Phi\,, \qquad \Phi_z=V\,\Phi\,,\label{LaxPair}
\end{equation}
where $U$ and $V$ are
\renewcommand\theequation{A2}
\begin{equation}\begin{array}{l}
\Phi=\left(\begin{array}{cc}
\varphi\\
\psi
\end{array}\right),\qquad
 U=-i\,\lambda\,J+\sqrt{\frac{R(z)}{d_{2}(z)}}\,\Lambda\\
 V=i\,4\,d_{3}(z)\lambda^{3}J-i\,\lambda^{2}d_{2}(z)\,J-4\lambda^{2}\,d_{3}(z)\sqrt{\frac{R(z)}{d_{2}(z)}}\,\Lambda+\lambda\, R(z)+\frac{i}{2}\,Q
\end{array}\end{equation}
with
\begin{equation*}\begin{array}{l}
J=\left(\begin{array}{cc}
1&0\\
0&-1
\end{array}\right),\qquad
\Lambda=\left(\begin{array}{cc}
0 & q \\
 -q^{*} &0 \\
\end{array}\right),\qquad\\
R=\left(\begin{array}{cc}
R_{11} & R_{12} \\
R_{21} & R_{22} \\
\end{array}\right)\,,\\
R_{11}=\frac{-2\,i\,d_{3}(z)|q|^{2}R(z)}{d_{2}(z)},\,
R_{12}=\sqrt{\frac{R(z)}{d_{2}(z)}}(d_{2}(z)q-2\,i\,d_{3}(z)\frac{\partial q}{\partial t}),\\
R_{21}=-\sqrt{\frac{R(z)}{d_{2}(z)}}(d_{2}(z)q^{*}+2\,i\,d_{3}(z)\frac{\partial q^{*}}{\partial t}),\,
R_{22}=\frac{2\,i\,d_{3}(z)|q|^{2}R(z)}{d_{2}(z)},\\
Q=\left(\begin{array}{cc}
Q_{11} & Q_{12} \\
Q_{21} & Q_{22} \\
\end{array}\right)\,,\\
Q_{11}=R(z)|q|^{2}-\frac{i\,d_{3}(z)R(z)}{d_{2}(z)}(2\,q^{*}\frac{\partial q}{\partial t}-2 q\frac{\partial q^{*}}{\partial t}),\,\\
Q_{12}=\sqrt{\frac{R(z)}{d_{2}(z)}}(d_{2}(z)\frac{\partial q}{\partial t}-2\,i\,d_{3}(z)(\frac{2\,q\,|q|^{2}R(z)}{d_{2}(z)}+\frac{\partial^{2}q}{\partial t^{2}})),\\
Q_{21}=\sqrt{\frac{R(z)}{d_{2}(z)}}(d_{2}(z)\frac{\partial q^{*}}{\partial t}+2\,i\,d_{3}(z)(\frac{2\,q^{*}\,|q|^{2}R(z)}{d_{2}(z)}+\frac{\partial^{2}q^{*}}{\partial t^{2}})),\,\\
Q_{22}=-R(z)|q|^{2}+\frac{i\,d_{3}(z)R(z)}{d_{2}(z)}(2\,q^{*}\frac{\partial q}{\partial t}-2 q\frac{\partial q^{*}}{\partial t}),
\end{array}\end{equation*}
$\lambda$ is the spectral parameter and $\Phi$ is the eigenfunction. Through direct computations, it can be verified that the equation $U_{z}-V_{t}+[U,V]=0$ exactly yields Eqs.~(\ref{vcHirota}). \\
\hspace*{\parindent}By using the transformation $\Phi^{[n]}=T\,\Phi$,\,we obtain the new Lax pair $\Phi^{[n]}_t=U^{[n]}\,\Phi^{[n]}$,\  $U^{[n]}=(T_t+T\,U)T^{-1}$,\  $\Phi^{[n]}_z=V^{[n]}\,\Phi^{[n]}$,\  $V^{[n]}=(T_z+T\,V)T^{-1}$,\, where $T$ is a $2\times 2$
matrix determined by the above relations $U^{[n]}_{z}-V^{[n]}_{t}+[U^{[n]},V^{[n]}]=T(U_{z}-V_{t}+[U,V])T^{-1}$.\\
\hspace*{\parindent}This implies that, in order to keep Lax Pair~(\ref{LaxPair}) invariant under the transformation, it is crucial to seek a matrix $T$ such that
$U^{[n]}$ and $V^{[n]}$ have the same forms as those of $U$ and $V$. In addition, the old potentials $q$ are mapped into new ones
$q^{[n]}$.

Next, we shall construct the \emph{n}-fold vc-modified Darboux transformation of Eqs.~(\ref{vcHirota}). Hereby, we assume Matrix $T_{n}$ be the form of
\renewcommand\theequation{A3}
\begin{equation}
 T_{n}=T_{n}(\lambda;\lambda_{1},\lambda_{2},...,\lambda_{2n})= \sum^{n}_{l=0}\,M_{l}\,\lambda^{n-l} \label{GTn}\,,
\end{equation}
where Matrices $M_{l}$ ($l=0, 1,2,...,n-1$) are solved by Cramer's rule, $\lambda_{k}=\alpha_{k}+i\,\beta\,\,(k=1, 2,...,2n$) denote the  spectral parameters and $M_{n}$ is an identity matrix.

Solving the linear system
\renewcommand\theequation{A4}
\begin{equation}
\Phi_{k}^{[n]}=T_{n}(\lambda;\lambda_{1},\lambda_{2},...,\lambda_{2n-1},\lambda_{2n})|_{\lambda=\lambda_{k}}\Phi_{k}= \sum^{n}_{l=0}\,M_{l}\,\lambda^{n-l}_{k}\,\Phi_{k}=0 \,\,\,(k=1,2,...,2n)\,,
\end{equation}
where $\Phi_{k}=(\varphi_{k}, \psi_{k})^{T}$ are the solutions of Lax Pair~(\ref{LaxPair}), we can get the determinant representation of the $T_{n}$ as follows
\renewcommand\theequation{A5}
\begin{equation}
 T_{n}=T_{n}(\lambda;\lambda_{1},\lambda_{2},...,\lambda_{2n})=\,\left(
\begin{array}{ll}
 \frac{{(\Omega_{n})_{11}}}{\Delta_{n}}  & \frac{{(\Omega_{n})_{12}}}{\Delta_{n}} \\
 \frac{{(\Omega_{n})_{21}}}{\Delta_{n}} & \frac{{(\Omega_{n})_{22}}}{\Delta_{n}}
\end{array}\right)  \,,\label{T}
\end{equation}
with {\footnotesize\begin{equation} \Delta_{n}=\begin{vmatrix}
\varphi_{1} & \psi_{1}& \cdots&\lambda_{1}^{n-2} \varphi_{1} &\lambda_{1}^{n-2}\psi_{1}&\lambda_{1}^{n-1} \varphi_{1}&\lambda_{1}^{n-1} \psi_{1}\\
\varphi_{2} & \psi_{2}&\cdots& \lambda_{2}^{n-2} \varphi_{2} &\lambda_{2}^{n-2} \psi_{2}&\lambda_{2}^{n-1} \varphi_{2}&\lambda_{2}^{n-1} \psi_{2}\\
\vdots&\vdots&\vdots&\vdots&\vdots&\vdots&\vdots\\
\varphi_{2n} & \psi_{2n}&\cdots& \lambda_{2n}^{n-2} \varphi_{2n} &\lambda_{2n}^{n-2} \psi_{2n}&\lambda_{2n}^{n-1} \varphi_{2n}&\lambda_{2n}^{n-1} \psi_{2n}
\end{vmatrix}\,,\nonumber
\end{equation}}
{\footnotesize\begin{equation} {(\Omega_{n})_{11}}=\begin{vmatrix}
1 &0& \lambda & 0& \cdots &\lambda^{n-1} &0&\lambda^{n}\\
\varphi_{1}&\psi_{1}&\lambda_{1}\varphi_1 &\lambda_{1}\psi_1&\cdots &\lambda_{1}^{n-1} \varphi_{1}&\lambda_{1}^{n-1}\psi_{1}&\lambda_{1}^{n}\varphi_{1}\\
\varphi_{2}&\psi_{2}&\lambda_{2}\varphi_2 &\lambda_{2}\psi_2&\cdots &\lambda_{2}^{n-1} \varphi_{2}&\lambda_{2}^{n-1}\psi_{2}&\lambda_{2}^{n}\varphi_{2}\\
\vdots&\vdots&\vdots&\vdots&\vdots&\vdots&\vdots&\vdots\\
\varphi_{2n}&\psi_{2n}&\lambda_{2n}\varphi_{2n} &\lambda_{2n}\psi_{2n}&\cdots &\lambda_{2n}^{n-1} \varphi_{2n}&\lambda_{2n}^{n-1}\psi_{2n}&\lambda_{2n}^{n}\varphi_{2n}
\end{vmatrix}\,,\nonumber
\end{equation}}
{\footnotesize\begin{equation} {(\Omega_{n})_{12}}=\begin{vmatrix}
0 &1& 0 & \lambda& \cdots &0&\lambda^{n-1}&0\\
\varphi_{1}&\psi_{1}&\lambda_{1}\varphi_1 &\lambda_{1}\psi_1&\cdots&\lambda_{1}^{n-1} \varphi_{1}&\lambda_{1}^{n-1}\psi_{1}&\lambda_{1}^{n}\varphi_{1}\\
\varphi_{2}&\psi_{2}&\lambda_{2}\varphi_2 &\lambda_{2}\psi_2&\cdots&\lambda_{2}^{n-1} \varphi_{2}&\lambda_{2}^{n-1}\psi_{2}&\lambda_{2}^{n}\varphi_{2}\\
\vdots&\vdots&\vdots&\vdots&\vdots&\vdots&\vdots&\vdots\\
\varphi_{2n}&\psi_{2n}&\lambda_{2n}\varphi_{2n} &\lambda_{2n}\psi_{2n}&\cdots&\lambda_{2n}^{n-1} \varphi_{2n}&\lambda_{2n}^{n-1}\psi_{2n}&\lambda_{2n}^{n}\varphi_{2n}
\end{vmatrix}\,,\nonumber
\end{equation}}

{\footnotesize\begin{equation}{(\Omega_{n})_{21}}=\begin{vmatrix}
1 &0& \lambda & 0& \cdots &\lambda^{n-1} &0&0\\
\varphi_{1}&\psi_{1}&\lambda_{1}\varphi_1 &\lambda_{1}\psi_1&\cdots&\lambda_{1}^{n-1} \varphi_{1}&\lambda_{1}^{n-1}\psi_{1}&\lambda_{1}^{n}\psi_{1}\\
\varphi_{2}&\psi_{2}&\lambda_{2}\varphi_2 &\lambda_{2}\psi_2&\cdots&\lambda_{2}^{n-1} \varphi_{2}&\lambda_{2}^{n-1}\psi_{2}&\lambda_{2}^{n}\psi_{2}\\
\vdots&\vdots&\vdots&\vdots&\vdots&\vdots&\vdots&\vdots\\
\varphi_{2n}&\psi_{2n}&\lambda_{2n}\varphi_{2n} &\lambda_{2n}\psi_{2n}&\cdots&\lambda_{2n}^{n-1} \varphi_{2n}&\lambda_{2n}^{n-1}\psi_{2n}&\lambda_{2n}^{n}\psi_{2n}
\end{vmatrix}\,,\nonumber
\end{equation}}
{\footnotesize\begin{equation} {(\Omega_{n})_{22}}=\begin{vmatrix}
0 &1& 0 & \lambda & \cdots &0 &\lambda^{n-1}&\lambda^{n}\\
\varphi_{1}&\psi_{1}&\lambda_{1}\varphi_1 &\lambda_{1}\psi_1&\cdots &\lambda_{1}^{n-1} \varphi_{1}&\lambda_{1}^{n-1}\psi_{1}&\lambda_{1}^{n}\psi_{1}\\
\varphi_{2}&\psi_{2}&\lambda_{2}\varphi_2 &\lambda_{2}\psi_2&\cdots &\lambda_{2}^{n-1} \varphi_{2}&\lambda_{2}^{n-1}\psi_{2}&\lambda_{2}^{n}\psi_{2}\\
\vdots&\vdots&\vdots&\vdots&\vdots&\vdots&\vdots&\vdots\\
\varphi_{2n}&\psi_{2n}&\lambda_{2n}\varphi_{2n} &\lambda_{2n}\psi_{2n}&\cdots&\lambda_{2n}^{n-1} \varphi_{2n}&\lambda_{2n}^{n-1}\psi_{2n}&\lambda_{2n}^{n}\psi_{2n}
\end{vmatrix}\,.\nonumber
\end{equation}}

For the \emph{n}-fold vc-modified Darboux transformation, the transformed potentials are
\renewcommand\theequation{A6}
{\footnotesize\begin{equation}
\begin{array}{l}
\Lambda^{[n]}=\Lambda+[J ,T_n]\,,\\
\end{array}
\end{equation}}
which produce the following \emph{n}-order solutions
\renewcommand\theequation{A7}
{\footnotesize\begin{gather} \label{ndt}
q^{[n]}=q^{[0]}-2\,i\,\frac{(M_1)_{12}}{\sqrt{R(z)/d_{2}(z)}}\,.
\end{gather}\label{n-order-solution}}
Note that
\renewcommand\theequation{A8}
\begin{equation}
\lambda_{2k}=-\lambda_{2k-1}^*\,,\qquad
\Phi_{2k}=\left(\begin{array}{cc}
\varphi_{2k}\\
\psi_{2k}
\end{array}\right)=
\left(\begin{array}{cc}
-\psi_{2k-1}^*\\
\varphi_{2k-1}^*
\end{array}\right)\,,\label{lambdann-1}
\end{equation}
in order to hold the constraints of \emph{n}-fold vc-modified Darboux transformation.

\bibliography{abanov-bibliography}

\end{document}